\documentclass[12pt,preprint]{aastex}

\newcommand{\etal}{{et~al.\,}}
\newcommand{\mic}{$\mu$m$\,$}
\newcommand{\teff}{T$_{\rm eff}\,$}

\begin{document}
\slugcomment{Submitted to Ap.J.}

\title{Spectroscopic Constants, Abundances, and Opacities of the TiH Molecule}

\author{A. Burrows\altaffilmark{1},
M. Dulick\altaffilmark{2}, C.W. Bauschlicher, Jr.\altaffilmark{3},
P.F. Bernath\altaffilmark{4,5}, R.S. Ram\altaffilmark{4},
C.M. Sharp\altaffilmark{1}, J.A. Milsom\altaffilmark{6}}

\altaffiltext{1}{Department of Astronomy and Steward Observatory, The University of Arizona,
Tucson, AZ \ 85721; burrows@zenith.as.arizona.edu, csharp@as.arizona.edu}
\altaffiltext{2}{National Solar Observatory, 950 N. Cherry Ave., P.O. Box 26732,
Tucson, AZ \ 85726-6732; dulick@noao.edu}
\altaffiltext{3}{NASA Ames Research Center, Mailstop 230-3, Moffett Field, CA \ 94035;
charles.w.bauschlicher@nasa.gov}
\altaffiltext{4}{Department of Chemistry, The University of Arizona, Tucson, AZ
\ 85721; rram@u.arizona.edu}
\altaffiltext{5}{Department of Chemistry, University of Waterloo, Waterloo, Ontario, Canada
N2L 3G1; bernath@uwaterloo.ca}
\altaffiltext{6}{Department of Physics, The University of Arizona,
Tucson, AZ \ 85721; milsom@physics.arizona.edu}

\begin{abstract}

Using previous measurements and quantum chemical calculations to
derive the molecular properties of the TiH molecule, we obtain new
values for its ro-vibrational constants, thermochemical data,
spectral line lists, line strengths, and absorption opacities.
Furthermore, we calculate the abundance of TiH in M and L dwarf
atmospheres and conclude that it is much higher than previously
thought. We find that the TiH/TiO ratio increases strongly with
decreasing metallicity, and at high temperatures can exceed unity.
We suggest that, particularly for subdwarf L and M dwarfs, spectral
features of TiH near $\sim$0.52 \mic, 0.94 \mic, and in the $H$ band
may be more easily measurable than heretofore thought.  The recent
possible identification in the L subdwarf 2MASS J0532 of the 0.94
\mic feature of TiH is in keeping with this expectation.  We
speculate that looking for TiH in other dwarfs and subdwarfs will
shed light on the distinctive titanium chemistry of the atmospheres
of substellar-mass objects and the dimmest stars.

\end{abstract}


\section{INTRODUCTION}
\label{intro}

The discovery of the new spectroscopic classes L and T (Nakajima et al. 1995; Oppenheimer et al. 1995;
Burgasser et al. 2000abc; Kirkpatrick et al. 1999) at lower effective
temperatures, (\teff), (2200 K$\rightarrow$700 K) than those
encountered in M dwarfs has necessitated the calculation (and recalculation) of
thermochemical and spectroscopic data for the molecules that predominate in such cool atmospheres.
At temperatures ($T$) below 2500 K, a variety of molecules not dominant in traditional
stellar atmospheres become important.  Though a few molecules, such as TiO, VO,
CO, and H$_2$O, have been features of M dwarf studies for some time,
for the study of L and T dwarfs, even for solar
metallicities, the molecules CH$_4$, FeH, CrH, CaH, and MgH
take on increasingly important roles.  The last four metal hydrides
are particularly important in subdwarfs, for which the metallicity is significantly
sub-solar.  In subdwarf M dwarfs, the spectral signatures of ``bimetallic" species such as TiO and VO weaken,
while those of the ``monometallic" hydrides, such as FeH, CaH, CrH, and MgH, increase in relative strength.
This shift from metal oxides to metal hydrides in subdwarfs is an old story for M dwarfs,
but recently, Burgasser et al. (2003) and Burgasser, Kirkpatrick \& L\'epine (2004)
have discovered two L dwarf subdwarfs, 2MASS J05325346+8246465 and 2MASS J16262034+3925190,
and this shift to hydrides is clear in them as well.  In fact, Burgasser, Kirkpatrick \& L\'epine (2004)
have recently tentatively identified titanium hydride (TiH) in
2MASS J0532 at $\sim$0.94 \mic, a molecule that heretofore
had not been seen in ``stellar" atmospheres (see also Burgasser 2004).

In support of L and T dwarf studies, we have an ongoing project to calculate
line lists, line strengths, abundances, and opacities of metal hydrides.  Burrows et al. (2002)
and Dulick et al. (2003) have already published new calculations for CrH and FeH, respectively.
With this paper, and in response to the recent detection of TiH in 2MASS J0532,
we add to this list new spectroscopic, thermochemical, and absorption opacity calculations for TiH.
In \S\ref{spectwork}, we summarize the spectroscopic measurements we use to help
constrain the spectroscopic constants of TiH.  We continue in \S\ref{methods} with
a discussion of the computational chemistry calculations we performed in conjunction
with our analysis of the laboratory work on TiH.
In \S\ref{EOS}, we calculate TiH abundances with the newly-derived thermochemical data
and partition functions and find that the TiH abundances, while not high,
are much higher than previous estimates.  We also describe the updates to our general
chemical abundance code that are most relevant to titanium chemistry.
Section \ref{opacity} describes how we use these lists to derive opacities
at any temperature and pressure and \S\ref{conclusions} summarizes our general conclusions
concerning these updated TiH ro-vibrational constants, abundances, and opacities.

\section{SUMMARY OF SPECTROSCOPIC WORK}
\label{spectwork}

Although spectra of TiH have been available for several decades (Smith \& Gaydon 1971), major
progress on the spectral assignment has been made only recently \citep{sssvh91,ll96,and03a,abblr2003}. The
spectrum of TiH was first observed by Smith and Gaydon in 1971 in a shock
tube and a tentative $^4\Delta-^4\Phi$ assignment was proposed for a band near 530 nm
(Smith \& Gaydon 1971). By comparing the TiH measurements (Smith
\& Gaydon 1971) with the spectra of $\alpha$ Ori, $\alpha$ Sco,
and $\delta$ Vir, Yerle (1979) proposed that TiH was present in stellar photospheres.
Yerle's tentative TiH assignments in these complex stellar spectra are very
dubious.

The first high-resolution investigation of TiH was made by Steimle
et al. (1991) who observed the laser excitation spectrum of the
530-nm band. After rotational analysis, this band was assigned as
the $^4\Gamma_{5/2}-X^4\Phi_{3/2}$ sub-band of a $^4\Gamma-X^4\Phi$
transition (called the $B^4\Gamma-X^4\Phi$ transition in the present
paper). The spectrum of the 530-nm transition of TiH was later
investigated by Launila and Lindgren (1996) by heating titanium
metal in an atmosphere of about 250 Torr of hydrogen, and the
spectra were recorded using a Fourier transform spectrometer. The
assignment of the 530-nm system was confirmed as a
$^4\Gamma-X^4\Phi$ transition by the rotational analysis of the four
sub-bands of the 0-0 vibrational band. The spectrum of the analogous
$^4\Gamma-X^4\Phi$ transition of TiD has recently been studied at
high resolution using a Fourier transform spectrometer
\citep{and03a}. TiD was also produced by heating titanium powder and
250 Torr of D$_2$ in a King furnace. The rotational analysis of the
$^4\Gamma_{7/2}-X^4\Phi_{5/2}$ and $^4\Gamma_{5/2}-X^4\Phi_{3/2}$
sub-bands produced the first rotational constants for TiD
\citep{and03a}.

In a more recent study, the spectra of TiH and TiD were observed in
the near infrared \citep{abblr2003} using the Fourier transform
spectrometer of the National Solar Observatory at the Kitt Peak. The
molecules were produced in a titanium hollow cathode lamp by
discharging a mixture of Ne and H$_2$ or D$_2$ gases. A new
transition with complex rotational structure was observed near 938
nm and was assigned as a $^4\Phi-X^4\Phi$ transition (called the
$A^4\Phi-X^4\Phi$ transition in the present paper). The complexity
of this transition is due to the presence of perturbations, as well
as overlapping from another unassigned transition \citep{abblr2003}.
The spectroscopic constants for the TiH and TiD states were obtained
by rotational analysis of the 0-0 band of the four sub-bands.

In other experimental studies of TiH, a dissociation energy of 48.9$\pm$2.1
kcal mole$^{-1}$ was obtained by Chen et al. (1991) using guided ion-beam mass
spectrometry. The fundamental vibrational intervals of 1385.3 cm$^{-1}$ for TiH
and 1003.6 cm$^{-1}$ for TiD were measured from matrix infrared absorption
spectra \citep{ca94}. The molecules were formed by the reaction of laser-ablated Ti atoms
with H$_2$ or D$_2$ and were isolated in an argon matrix at 10 K.

There have been several previous theoretical investigations of TiH. Ab
initio calculations for TiH and other first-row transition metal hydrides
have been carried out by Walch and Bauschlicher (1983), Chong et al. (1986), and
Anglada et al. (1990). Spectroscopic properties of the ground states of the
first row transition metal hydrides were predicted in these studies. Chong
et al. (1986) and Anglada et al. (1990) have also calculated spectroscopic
properties of some low-lying states. In a recent study the spectroscopic
properties of the ground and some low-lying electronic states of TiH were
computed by Koseki et al. (2002) by high level ab initio calculations that
included spin-orbit coupling. The potential energy curves of low-lying states
were calculated using both effective core potentials and all-electron
approaches.

\section{COMPUTATIONAL CHEMISTRY METHODS AND RESULTS}
\label{methods}

To supplement the sparse spectroscopic measurements summarized in \S\ref{spectwork},
we calculate the spectroscopic constants, transition dipole moments, line
strengths, and ro-vibrational constants of the TiH molecule using
standard ab initio quantum chemical techniques.
The orbitals of the TiH molecule are optimized using a state-averaged
complete-active-space self-consistent-field (SA-CASSCF) approach (Roos, Taylor \& Siegbahn 1980),
with symmetry and equivalence restrictions imposed.  Our
initial choice for the active space include the Ti 3d, 4s, and 4p orbitals and the H
1s orbital. In $C_{2v}$ symmetry, this active space corresponds to four $a_1$, two $b_1$,  two $b_2$,
and one $a_2$ orbital, which is denoted (4221).  Test calculations showed that this active
space is much too small to study both the $B{}^4\Gamma-{}X^4\Phi$ and $A{}^4\Phi-{}X^4\Phi$
systems simultaneously.  In fact, no practical choice for the active space was found
that allowed the study of both band systems simultaneously.
Therefore, the two band systems are studied using separate calculations.  For
the $B{}^4\Gamma-{}X^4\Phi$ system, the (4221) active space  is sufficient.
In this  series  of calculations, two $^4\Delta$ states and one state each of $^4\Pi$,
$^4\Phi$, and $^4\Gamma$ are included in the SA procedure, with
equal weight given to each state.  In the $A{}^4\Phi-{}X^4\Phi$ SA-CASSCF calculations, it was necessary
to  increase the active space, and our final choice is a (7222) active space that has two
additional $\sigma$ orbitals  and one additional $\delta$ orbital, relative to the (4221)
active space.  Only the two $^4\Phi$ states are included in the SA-CASSCF calculations for the
$A{}^4\Phi-{}X^4\Phi$ system.

More extensive correlation is included using the internally
contracted multireference configuration interaction
(IC-MRCI) approach~\citep{ic1,ic2}.  All configurations from the CASSCF calculations
are used as references in the IC-MRCI calculations.  The valence calculations
correlate the  Ti 3d and 4s  electrons and the H electron, while in the core+valence
calculations the Ti 3s and 3p electrons are also correlated.  The Ti 3s- and 3p-like orbitals are in the
inactive space so they are doubly occupied in all reference configurations.
The effect of higher excitations is accounted for using the multireference analogue of
the Davidson correction (+Q) (Langhoff \& Davidson 1974).  Since the IC-MRCI calculations are
performed in $C_{2v}$ symmetry, the $\Delta$
and $\Gamma$ states are in the same symmetry as are the $\Phi$ and $\Pi$ states.
Thus, for the $B{}^4\Gamma-{}X^4\Phi$ calculation, the IC-MRCI reference
space included three roots of $^4A_1$ symmetry
and two roots of $^4B_1$ symmetry.
For the $A{}^4\Phi-{}X^4\Phi$ system, the reference space
includes four roots of $^4B_1$ symmetry since there are
two $^4\Pi$ states below the $A{}^4\Phi$ state.

Scalar relativistic effects are accounted for using
the Douglas-Kroll-Hess (DKH) approach \citep{hess}.
For Ti we use the (21s16p9d4f3g1h)/[7s8p6d4f3g1h]
quadruple zeta (QZ) 3s3p basis set~\citep{tiqz}, while
for H we use the correlation-consistent polarized
valence QZ set~\citep{cc1}.
The nonrelativistic contraction coefficients are replaced
with those from DKH calculations.  All calculations
are performed with Molpro2002~\citep{molpro} that
is modified to compute the DKH integrals~[C.W. Bauschlicher, unpublished].

The transition dipole moments for the forbidden $^4\Pi$-$^4\Gamma$ and $^4\Pi$-$^4\Phi$
transitions are close to zero. Therefore, we conclude
that the states are quite pure despite the
fact that the IC-MRCI calculations are performed in
$C_{2v}$ symmetry.  We summarize the computed spectroscopic
constants along with available experimental
data \cite{abblr2003,sssvh91,ll96,ca94} in Table~\ref{theory1}.
We do not include the (2)$^4\Delta$ state from
the $B{}^4\Gamma-X{}^4\Phi$ calculation, since, based on the
weight of the reference configurations in IC-MRCI calculations,
this state is not well described by these
calculations.

We first compare the results for the $X{}^4\Phi$
state as a function of treatment.  For the treatments that include only
valence correlation, $A{}^4\Phi-X{}^4\Phi$ and $B{}^4\Gamma-X{}^4\Phi$ calculations yield
very similar results for the $X{}^4\Phi$ state.  When
core correlation is included, there is a larger difference
between the two sets of calculations. For example, note
that the $\omega_e$ values (see Table \ref{theory1}) which
differ by 6.4~cm$^{-1}$ at the valence level, differ
by 19.7~cm$^{-1}$ when core correlation is included.  Our
computed $\omega_e$ values are significantly larger than
the value assigned by Chertihin and Andrews (1994)
to TiH in an Ar matrix. However, recently
L. Andrews~[personal communication] has suggested that their value is incorrect.
The $A{}^4\Phi-X{}^4\Phi$ core+valence treatment
results in a small increase in the dipole moment, which
is now in excellent agreement with experiment, while for
the $B{}^4\Gamma-X{}^4\Phi$ treatment, the inclusion of
core correlation leads to a significant decrease in
size and a value that is significantly smaller than experiment.
Using a finite field approach for the $X$ state in the $B{}^4\Gamma-X{}^4\Phi$
core+valence treatment yields a value that is in better
agreement with experiment, but is still 0.29~D too small.
The $A{}^4\Phi-X{}^4\Phi$ core+valence treatment
yields the best agreement with experiment for the
value of the Ti-H bond distance ($r_0$) of the $X$ state,
in addition to yielding the best dipole moment.

The core+valence treatment yields a better $r_0$ value
for the $A{}^4\Phi$ state, but the disagreement
with experiment is larger than for the $X$ state.
The valence treatment yields a somewhat better term value ($T_0$),
with both the valence and core+valence values being
larger than experiment.  The inclusion of core correlation
improves the agreement with experiment for the $r_0$ value for the $B{}^4\Gamma$ state and
also improves the agreement with the experimental $T_0$.
While the change in most $\omega_e$ values with the
inclusion of core correlation is small, the inclusion
of core correlation leads to a sizable reduction in $\omega_e$ for the $A{}^4\Phi$ state.
We observe a decrease in the dipole moments of
the $^4\Pi$, $^4\Delta$, and $B{}^4\Gamma$ states
when core correlation is included, as found for
the $X$ state in the $B{}^4\Gamma-X{}^4\Phi$ treatment.
As for the $X$ state, the valence level dipole moment for the $B{}^4\Gamma$
state agrees better with experiment.  We suspect
that true values for the $^4\Pi$ and $^4\Delta$ states
are closer to those obtained in the valence treatment
than those obtained in the core+valence treatment.

While the dipole moments vary significantly when core correlation is included,
the effect of including core correlation on the
dipole transition moment is only up to about 3~percent in the
Franck-Condon region for both transitions of
interest.  For the $A{}^4\Phi-X{}^4\Phi$ transition,
core correlation increases the moment, while for
the $B{}^4\Gamma-X{}^4\Phi$ transition core correlation decreases
the moment.

\begin{figure}
\plotone{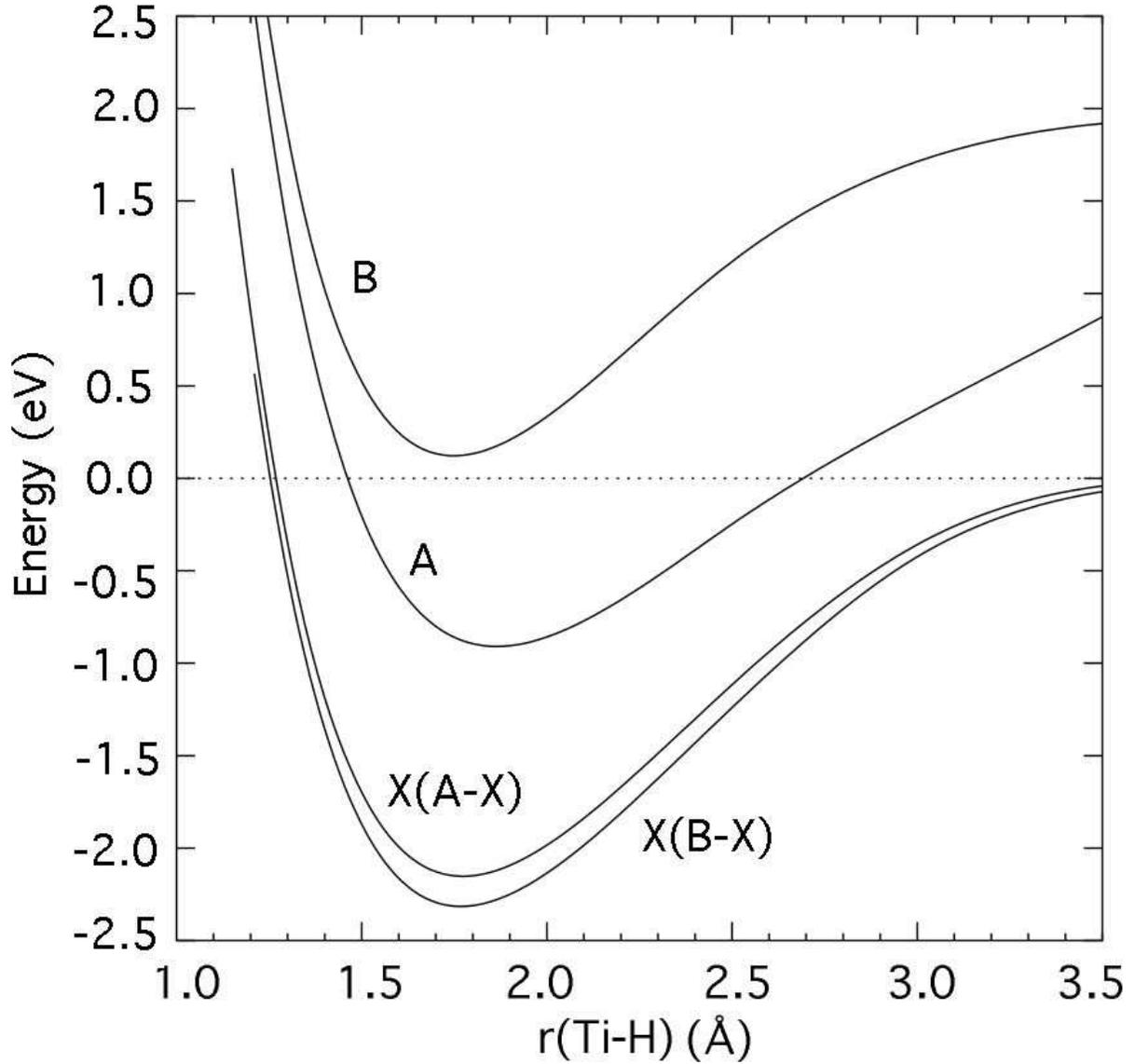} \caption{ \label{fig:t1}The core+valence IC-MRCI+Q
potential energy curves for the $X$, $A$, and $B$ states. The zero
of energy is Ti and H at infinite separation.  The $X$ state is
plotted twice, one curve is from the $A-X$ calculations while the
second is from the B-X calculations.  See the text for a full
description of the two sets of calculations. }
\end{figure}

\begin{figure}
\plotone{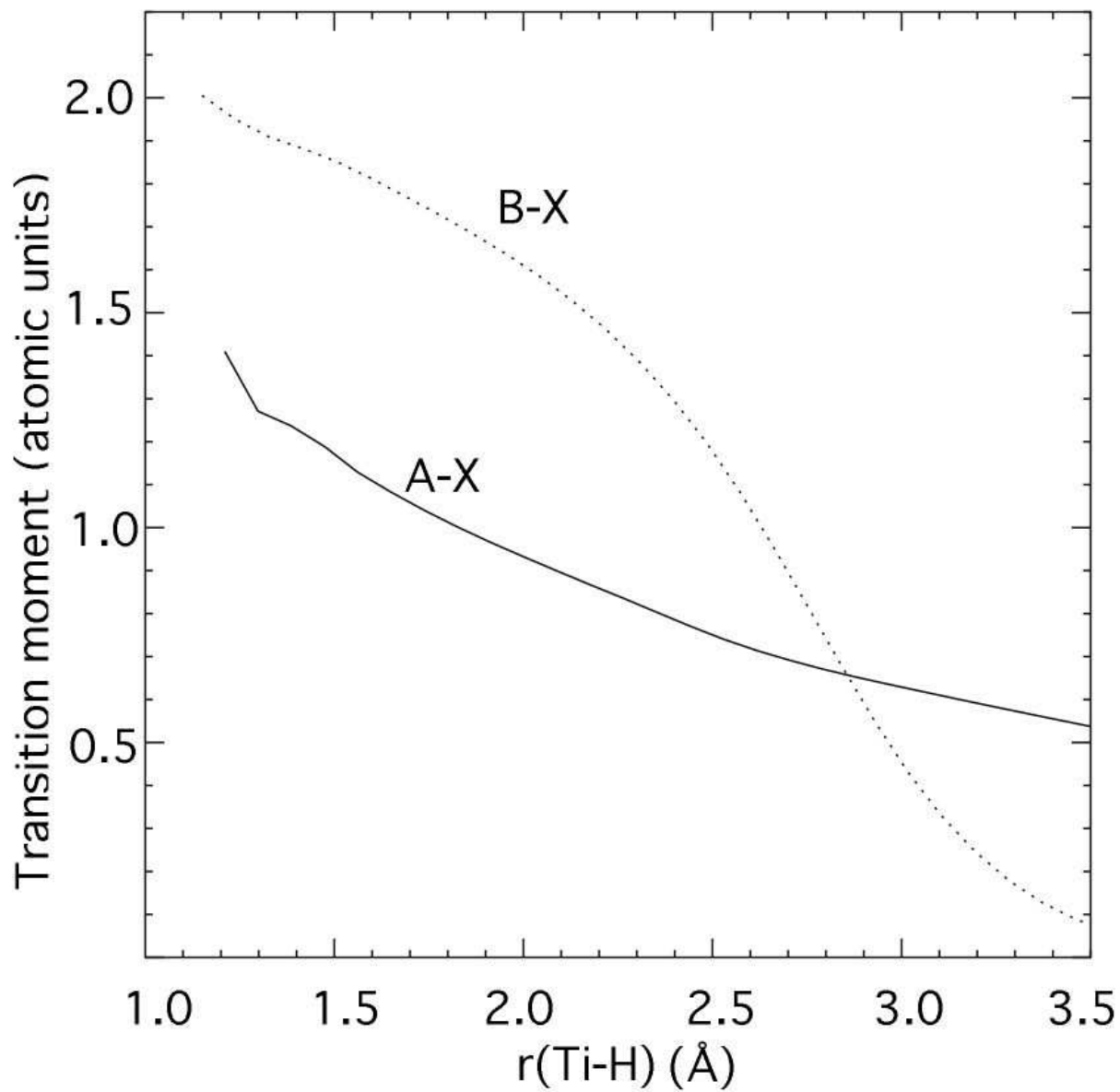}
\caption{ \label{fig:t2}
 The core+valence IC-MRCI $A-X$ and $B-X$ transition dipole moments. }
\end{figure}

Given the limited experimental data it is difficult to definitively
determine the best values for $\omega_e$, $\omega_e x_e$, and the
transition moment.  On the basis of previous calculation on CrH
\cite{crh,bur} and FeH \cite{dulick}, we feel that the calculations
including the core+valence may yield superior values for all
spectroscopic constants excluding the dipole moment, where the
smaller active space used in the $B{}^4\Gamma-X{}^4\Phi$
calculations results in the valence treatment being superior for
this property.  For the $X{}^4\Phi$ state, we believe the results
obtained using the $A{}^4\Phi-X{}^4\Phi$ treatment are superior
because it has a larger active space and fewer states in the SA
procedure. We also note that for the $A{}^4\Phi-X{}^4\Phi$
treatment, the $X$ state dipole moment improves when core
correlation is included, which also supports using the
$A{}^4\Phi-X{}^4\Phi$ treatment for the $X$ state. In the end,
transition dipole moment functions computed at the core+valence
level were used to generate the Einstein $A_{v^\prime
v^{\prime\prime}}$ values for the $B{}^4\Gamma-X{}^4\Phi$ and
$A{}^4\Phi-X{}^4\Phi$ transitions in Table 2. The transition moments
and IC-MRCI+Q potentials used to compute the $A$ values in Table 2
are plotted in Figures \ref{fig:t1} and \ref{fig:t2},
respectively.\par
  Finally, we compare our best values to previous theoretical values.
Our best results are in reasonable agreement with those reported by
Anglada et al. (1990); for the $X$ and $A$ states their bond
distances are 0.07 and 0.02~\AA\ longer than our best values,
respectively, and therefore in slightly worse agreement with
experiment.  Their $X$ state $\omega_e$ value is about 50~cm$^{-1}$
smaller than our best value, while their $A$ state value is about
140~cm$^{-1}$ larger than our best value.  For the one bond distance
for which they report their transition moments, their $A-X$
transition moment differs from our value by about 10\%, while their
$B-X$ moment is only very slightly smaller than our value.  The $X$
state $r_e$ values reported by Koseki et al. (2002) are about
0.1~\AA\ longer than our $r_e$ value.  Their all electron $\omega_e$
value is in better agreement with our value than is their value with
scalar relativistic effects added.  This is a bit surprising since
our calculations include the scalar relativistic effects. Overall
the agreement between the different theoretical approaches is
relatively good.  Our use of much larger basis sets and the
inclusion of core-valence correlation results in bond lengths that
are in better agreement with experiment, and we assume that the
other spectroscopic properties are more accurate in our calculations
as well.

\section{LINE POSITIONS AND INTENSITIES}
\label{dulick}

A common feature associated with the electronic structure of transition
metal hydrides is that the excited electronic states are often heavily
perturbed by unobserved neighboring states whereas the ground states are
for the most part unperturbed. The $X^4\Phi$, $A^4\Phi$, and $B^4\Gamma$
states of TiH are no exception. Recent work on the analyses of
high-resolution Fourier transform emission and laser excitation spectra
of the $A^4\Phi - X^4\Phi$ (Andersson et al. 2003) and $B^4\Gamma - X^4\Phi$
(Steimle et al. 1991; Launila \& Lindgren 1996) $0-0$ bands have finally
yielded rotational assignments for the heavily perturbed $A$ and $B$ states.

As a prerequisite to calculating opacities for temperatures up to
2000 K, a synthesized spectrum was generated comprised of
rovibrational line positions and Einstein $A$-values for quantum
numbers up to $v^{\prime\prime}=v^\prime=5$ and
$J^{\prime\prime}=J^\prime=50.5$. Because the only available
experimental data for these states is for $v=0$, we had to rely
heavily on the ab initio calculations described above to supply the
missing information on the vibrational structure. In particular, the
first 4 vibrational levels were calculated from the ab initio
potentials and fitted with 3 vibrational constants as reported at
the end of Table 6. Note that the vibrational constants of Table 1
were generated in the same way but with 3 vibrational levels and 2
fitted vibrational constants (as listed in Table 1). The vibrational
constants in Tables 1 and 6 are therefore slightly different.

The electronic coupling in the $X$, $A$, and $B$ states is
intermediate Hund's case (a) -- (b) over a wide range of $J$. The
case (a) Hamiltonians for the $^4\Phi$ and $^4\Gamma$ states in
Tables 3 and 4, respectively, were selected for least-squares
fitting of the experimental data. In determining the molecular
constants for the $X$ state, the reported rotational lines in
Andersson et al. (2003) were reduced to lower state combination
differences, yielding the molecular constants listed in Table 5.
>From these fitted constants, the $X$ state term energies were
calculated and then used in transforming the observed rotational
lines in the $A-X$ and $B-X$ $0-0$ bands into term energies for the
$A$ and $B$ states. Molecular constants derived from fits of the $A$
and $B$ term energies are also listed in Table 5.

Except for the $X$ state, the centrifugal constants $D_0$ and $H_0$ were
not statistically determined. Instead, these constants were held fixed in
the fits of the $A$ and $B$ states to the values determined by the
Dunham approximations,
\begin{eqnarray}
D_v &\simeq & \xi^2 B_v\hspace{10pt}\mbox{and}\hspace{10pt}
H_v \:\simeq\: \xi^5 \omega_e - \xi^3 \alpha_e / 3
\end{eqnarray}
where $\xi = 2 B_v / \omega_e$ and estimated values for $\omega_e$
and $\alpha_e$ were supplied by our ab initio calculations (Table 6).
The presence of an anomalously large $e/f$ parity splitting in the $A$ state,
undoubtedly due to perturbations, required separate fits of the $e$- and
$f$-parity term energies, yielding the two sets of molecular constants in
Table 5. Finally, the standard deviations from the fits, 0.031, 0.28,
0.16, and 1.18 cm$^{-1}$ for the $X$, $A$(e), $A$(f), and $B$ states, are
comparable in quality to 0.026 and 0.42 cm$^{-1}$ for $X$ and $A$
(Andersson et al. 2003) and 0.020 and 0.573 cm$^{-1}$ for $X$ and $B$
(Launila \& Lingren 1996) with roughly the same number of data points and
adjustable parameters.

The molecular constants for $v > 0$ were generated in a straightforward
manner. With the ab initio estimates, $\omega_e$, $\omega_e x_e$,
$\omega_e y_e$, $\alpha_e$, $\gamma_e$, and $\epsilon_e$, for each state
(Table 6), the rotational constants for $v > 0$ were computed from
\begin{eqnarray}
B_v &=& B_e - \alpha_e (v+1/2) + \gamma_e (v+1/2)^2 + \epsilon_e (v+1/2)^3
\end{eqnarray}
where the $B_e$'s were determined from $B_0$'s, the centrifugal
constants $D_v$ and $H_v$ from eq.\ (1), and the spin-component energies
$T_v(\Omega)$ by starting from the $T_0(\Omega)$ values and recursively
adding vibrational separations of
\begin{eqnarray}
\Delta G(v) &=& \omega_e - 2 v \:\omega_e x_e + (3 v^2 + 1/4)\:\omega_e y_e
\end{eqnarray}
for $v > 0$ in which the dependence of $A$, $\lambda$, and $\gamma$
on $v$ was assumed to be negligible. The list of molecular constants
in Table 6 were then used with the $^4\Phi$ and $^4\Gamma$
Hamiltonians in computing the term energies from $v=0$ to $v=5$ for
$e/f$ rotational levels up to $J=50.5$ for the $X$, $A(e)$, $A(f)$,
and $B$ states. Note that all of the vibrational constants have been
obtained using ab initio calculations so the location of bands with
$\Delta v \not= 0$ (computed using equation (3)) is relatively
uncertain.

In our calculations the Einstein $A$-value is partitioned as
\begin{eqnarray*}
A &=& A_{v^\prime v^{\prime\prime}}\:\mbox{HLF} / (2 J^\prime + 1)
\end{eqnarray*}
where $A_{v^\prime v^{\prime\prime}}$ is the Einstein $A$ for the
vibronic transition $v^{\prime} - v^{\prime\prime}$, calculated
using the ab initio electronic transition dipole moment function,
and HLF is the H\"onl--London factor. For reference, the
H\"onl-London factors (labeled as $S(J)$) in Tables 7 -- 10 for
$^4\Phi - {^4\Phi}$ and $^4\Gamma - {^4\Phi}$ transitions involving
pure Hund's case (a) and (b) coupling were derived using the method
discussed in Dulick et al. (2003). Because the coupling in the $X$,
$A$, and $B$ states is intermediate Hund's case (a) -- (b), the
actual HLF's were computed by starting from
$\mbox{\bf{U}}^\dagger_l\mbox{\bf{T}}\mbox{\bf{U}}_u$, where
$\mbox{\bf{U}}_l$ and $\mbox{\bf{U}}_u$ are the eigenvectors
obtained from the diagonalizations of the lower and upper state
Hamiltonians and $\mbox{\bf{T}}$ is the electric-dipole transition
matrix (cf., eq.\ (1) in Dulick et al. (2003)). Squaring the matrix
elements of $\mbox{\bf{U}}^\dagger_l\mbox{\bf{T}}\mbox{\bf{U}}_u$
yield numbers proportional to the HLF's for intermediate coupling.
In the absence of perturbations from neighboring states, as $J$
becomes large, the intermediate case HLF's asymptotically approach
the pure Hund's case (b) values.

\section{CHEMICAL ABUNDANCES OF TITANIUM HYDRIDE}
\label{EOS}

In order to estimate the contribution ${\rm TiH}$ makes to M dwarf spectra,
its abundance must be calculated. This is accomplished by minimizing
the total free energy of the ensemble of species using the same methods as
Sharp \& Huebner (1990) and Burrows \& Sharp (1999) (see Appendix A).
To calculate the abundance of a particular species,
its Gibbs free energy of formation from its constituent elements must be known.
Using the new spectroscopic constants derived in \S\ref{spectwork} and \S\ref{methods},
we have calculated for TiH this free energy and its associated partition function.
In the case of ${\rm TiH}$, its dissociation into its constituent elements can be written as
the equilibrium
$${\rm TiH} \rightleftharpoons {\rm Ti} + {\rm H} \, . $$
The Gibbs energy of formation is then calculated from eq. (\ref{DATA}) in
Appendix A. Eq. (\ref{DATA}) is the general formula for obtaining the energy of
formation of a gas-phase atomic or molecular species, be it charged or
neutral.

Since the publication of Burrows \& Sharp (1999), we have upgraded
our chemical database significantly with the inclusion of several additional
titanium-bearing compounds, and have updated most of
the thermochemical data for titanium-bearing species (Barin 1995). In addition to ${\rm TiH}$,
the new species are the solid condensates ${\rm TiS}$, ${\rm Mg_2TiO_4}$,
${\rm Ca_3Ti_2O_7}$, and ${\rm Ca_4Ti_3O_{10}}$, and the revised data are for
the gas-phase species ${\rm TiO}$, TiO$_2$, and ${\rm TiS}$, the liquid condensates
${\rm TiN}$, ${\rm TiO_2}$, ${\rm Ti_2O_3}$, ${\rm Ti_3O_5}$, ${\rm Ti_4O_7}$,
${\rm MgTiO_3}$, ${\rm Mg_2TiO_4}$, and the solid condensates
${\rm TiN}$, ${\rm TiO_2}$, ${\rm Ti_2O_3}$, ${\rm Ti_3O_5}$, ${\rm Ti_4O_7}$,
${\rm MgTiO_3}$, ${\rm CaTiO_3}$, and ${\rm MgTi_2O_5}$.  The revised data for TiO$_2$,
in particular, have resulted in refined abundance estimates of TiO.  We have obtained much
better least-square fits to the free energy, and in nearly all cases the fitted
polynomials agree to better than 10 cal mole$^{-1}$ compared with the tabulated
values in Barin (1995) over the fitted temperature range, which for condensates
corresponds to their stability field. For computational efficiency we sometimes
extrapolate the polynomial fits of eq. (\ref{GT}) for the condensates beyond the temperature
range in which they are stable, but we ensure that the extrapolation is in a
direction that further decreases stability.

Since ionization plays a significant role at higher temperatures, a number of
ionized species are included in the gas phase, including electrons
as an additional ``element" with a negative stoichiometric coefficient.
The ions considered in our equilibrium code are ${\rm e^-}$,
${\rm H^+}$, ${\rm H^-}$, ${\rm Li^+}$, ${\rm Na^+}$, ${\rm K^+}$, ${\rm Cs^+}$,
${\rm Mg^+}$, ${\rm Al^+}$, ${\rm Si^+}$, ${\rm Ti^+}$, ${\rm Fe^+}$, ${\rm H_2^+}$,
${\rm H_2^-}$, ${\rm OH^+}$, ${\rm CO^+}$, ${\rm NO^+}$, ${\rm N_2^+}$,
${\rm N_2^-}$, and ${\rm H_3O^+}$. Ions were not considered in the earlier work
of Burrows and Sharp (1999).

The partition function of ${\rm TiH}$ is obtained from the estimated
spectroscopic constants in Tables \ref{table:X1} and \ref{table:X2}
(data taken from Table 1 and Anglada et al. 1990, or simply
guessed). Table \ref{table:X1} lists 15 electronic states, with the
quartet $X^4\Phi$ ground state being resolved into each of the four
separate spin substates. The partition function is calculated using
the method of Dulick et al. (2003). The contribution to the total
partition function of each electronic state is determined using
asymptotic approximations from the vibrational and rotational
constants in Table \ref{table:X2}, then these contributions are
summed according to eq. (7) in Dulick et al. (2003), with the
Boltzmann factor for the electronic energy being applied to each
state, where the electronic energy of the lowest spin substate of
the ground $X$ electronic state is zero by definition. Since the
separate electronic energies for the spin substates of the ground
electronic state are available, the $X$ state is treated as four
separate states in the summation, so that effectively the sum is
over 18 states. The partition function of titanium is obtained by
summing over the term values below 20,000 cm$^{-1}$, with
degeneracies as listed in Moore (1949). The partition function of
hydrogen is set to a value of 2 for all the temperatures considered.
Table \ref{table:X3} lists our resulting TiH partition function 
in steps of 200 K from 1200 K to 4800 K. 
Using the partition functions of ${\rm TiH}$, ${\rm Ti}$, and ${\rm
H}$ in eq. (\ref{DATA}), $\Delta G(T)$ is calculated at 100 K
intervals over the temperature range required to compute the
abundance of ${\rm TiH}$, fitted to a polynomial, and then
incorporated into our database.

\begin{figure}
\plotone{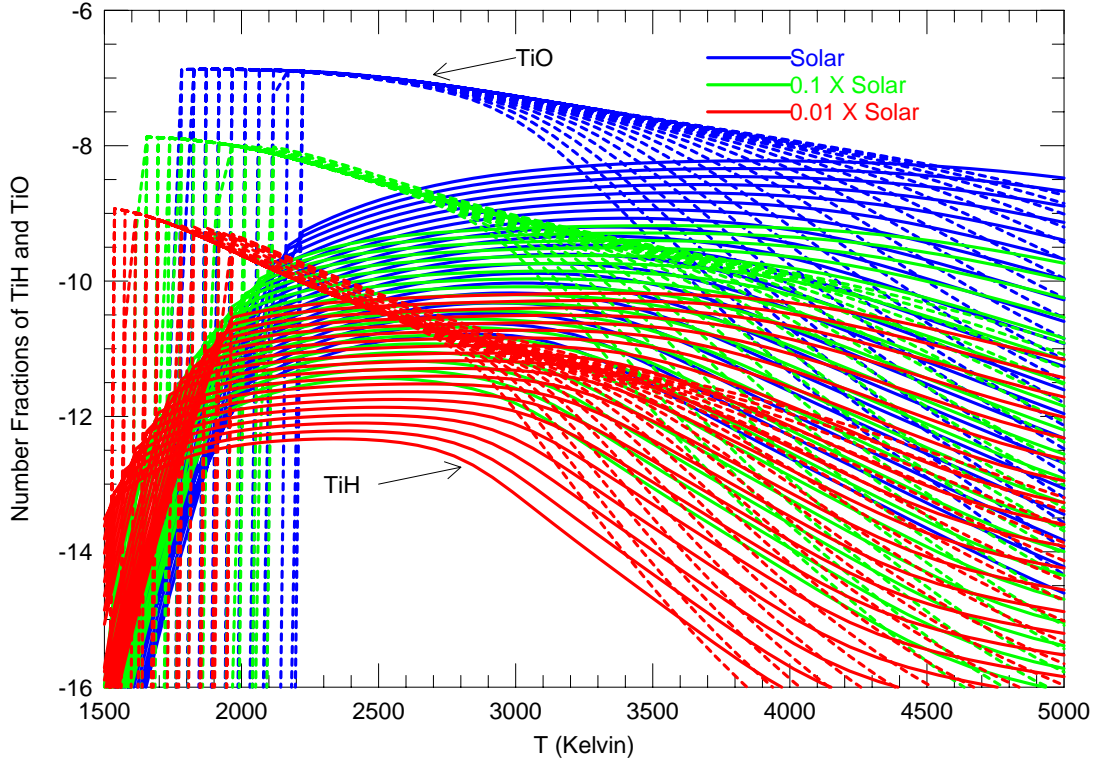}
\caption{ \label{fig:t3}
The log (base 10) of the number fractions of both TiH (solid) and TiO (dashed)
as a function of temperature (T) from 1500 to 5000 Kelvin, for pressures from 10$^{-2}$
atmospheres to 10$^{2}$ atmospheres in equal logarithmic steps.  Pressure increases 
for a given sequence from bottom to top. Furthermore, the condensation lines march
rightward as the pressure increases.  
The abundances of both species are given for solar metallicity
(blue), $0.1 \times$ solar metallicity (green), and $0.01 \times$ solar metallicity (red).
TiO generally dominates at lower temperatures, until titanium condensables,
such as perovskite, form, after which the abundances of all titanium
compounds decrease precipitously (left of diagram).  The condensation temperature
increases with increasing pressure.  However, at high temperatures,
TiH comes into its own and its abundance can exceed that of TiO.  For a given
pressure, the temperature above which the TiH/TiO ratio exceeds unity is an increasing
function of metallicity, being at 10$^2$ atmospheres and solar metallicity $\sim$4500 K,
but at 10$^2$ atmospheres and $0.01 \times$ solar metallicity $\sim$2500 K.  As metallicity decreases, TiH
becomes more and more important, though the abundances of both TiO and TiH decrease
with metallicity. When the first titanium-bearing condensate appears, the abundance of TiH
is already falling.  Note that with increasing pressure the peak in TiH abundance shifts
to higher temperatures more rapidly than the condensation temperature. }
\end{figure}

The thermochemistry indicates that the most important species of titanium
are ${\rm TiO}$, ${\rm TiO_2}$, ${\rm TiH}$, ${\rm Ti}$, and ${\rm Ti^+}$.
${\rm TiO_2}$ is always subdominant and ${\rm Ti}$ and ${\rm Ti^+}$
are important only at higher temperatures than obtain in M, L, or T dwarf atmospheres.
At solar metallicity, just before the first titanium-bearing condensate appears,
${\rm TiO_2}$ replaces Ti as the second most abundant
titanium-bearing species, reaching an abundance of $\sim$9\%.
Figure \ref{fig:t3} depicts the dependence of the TiO and TiH abundances with temperature
from 1500 K to 5000 K for a range of pressures ($P$) from 10$^{-2}$ atmospheres
to 10$^2$ atmospheres.  This figure also shows the metallicity dependence of
the TiO and TiH abundances.  For solar metallicity, TiO dominates at low pressures
below $\sim$4000 K and at high pressures below $\sim$4500 K, but the TiH
abundances are not small.  For solar metallicity,
the TiH/TiO ratio at 10$^{-2}$ atmospheres and 2500 K is 10$^{-4}$ and at
10$^2$ atmospheres and 2500 K it is $\sim$10$^{-2}$ (one percent).
The predominance of ${\rm TiO}$ before condensation, then its disappearance
with condensation, matches the behavior seen in the M to L dwarf transition.

However, the abundance of TiO is a more strongly decreasing function
of temperature than is that of TiH.  For a given pressure, there is
a temperature above which the TiH/TiO ratio actually goes above one.
As Fig. \ref{fig:t3} indicates,
this transition temperature decreases with decreasing metallicity.  For 0.01$\times$ solar
metallicity and a pressure of 1 atmosphere, the TiH/TiO ratio hits unity at only $\sim$3500 K
and at 10$^2$ atmospheres it does so near 2500 K.  As a result,
we expect that TiH will assume a more important role in low-metallicity subdwarfs
than in ``normal" dwarfs, with metallicities near solar.  This fact is in
keeping with the putative measurement of the TiH band near 0.94 \mic (\S\ref{opacity})
in the subdwarf 2MASS J0532.

As expected, the abundance
of gas-phase species containing titanium drops rapidly when the first
titanium-bearing condensates appear, which occurs for solar metallicity at temperatures
between 2200 K and 1700 K for pressures of 10$^2$ and 10$^{-2}$
atmospheres, respectively (see Fig. \ref{fig:t3}).
At solar metallicity, when
titanium-bearing condensates form, the TiH abundance is already
falling with decreasing temperature; at condensation its abundance is at
least an order of magnitude down from its peak abundance.
As Fig. \ref{fig:t3} indicates, the TiH abundance is an increasing function
of pressure, but assumes a peak, sometimes broad, between the formation of condensates
at lower temperatures and the formation of atomic Ti/Ti$^+$ at higher temperatures.

The only significant previous study that included TiH is that of
Hauschildt et al. (1997). These authors present numerical data in
the form of fractional abundances in the gas phase for a number of
titanium-bearing species, including TiO and TiH, as a function of
optical depth. Unfortunately, this makes comparison with our results
difficult.  Nevertheless, we can still see that for their two
solar-metallicity models the maximum ratio of the abundance of TiH
to TiO is smaller than we obtain, sometimes by large factors. For
instance, for their T$_{\rm eff}$=2700 K model, at an optical depth
of unity, their ${\rm TiH}$ abundance is over four orders of
magnitude lower than that of TiO, and at an optical depth of 100, it
is still down by three orders of magnitude. For their T$_{\rm
eff}$=4000 K model at unit optical depth, their ${\rm TiH}$
abundance is below that of TiO by three orders of magnitude.  These
very low abundances are in contrast to what we obtain at comparable
$T/P$ points (cf. Fig. \ref{fig:t3}).

On the other hand, we find at a temperature of 2700 K, a
pressure of 100 atmospheres, and solar metallicity
that the abundance of TiH is as high as $\sim$1.5\%
of that of TiO. At 1 atmosphere, this ratio has dropped by about a
factor of 10, but is still about a factor of 10 larger than the corresponding ratio
in Hauschildt et al.  At a temperature of 4000 K, a pressure of 100 atmospheres,
and solar metallicity, we obtain a TiH/TiO abundance ratio as large as $\sim$24\%.
Furthermore, as previously mentioned, we find that the TiH/TiO ratio can go above one
and at low metallicities is above one for a wider temperature range.
Even at temperatures where this ratio is significantly below one, we conclude that the TiH
abundance derived with the new thermochemical data is much higher
than previously estimated.

\section{TITANIUM HYDRIDE ABSORPTION OPACITIES}
\label{opacity}

Using the results of \S\ref{spectwork}, \S\ref{methods}, and
\S\ref{dulick}, we calculate the opacity of the two electronic
systems $A^4\Phi - X^4\Phi$ and $B^4\Gamma - X^4\Phi$ of ${\rm
TiH}$. These have band systems centered near 0.94 \mic in the near
IR and 0.53 \mic in the visible, respectively. For the $A-X$ system,
4556 vibration-rotation transitions are considered for each of the
20 vibrational bands $v'-v''$ with $v'$=0 to 4 and $v''$=0 to 3,
giving 91120 lines in total. For the $B-X$ system, 4498
vibration-rotation transitions are incorporated for each of the 24
vibrational bands $v'$=0 to 3 and $v''$=0 to 5, giving 107,952 lines
in total.

We use the same methods to calculate the line strengths and
broadening as in \S{5} of Dulick et al. (2003), using the partition
function of TiH from \S\ref{EOS}. Since broadening parameters for
TiH are not available, those of ${\rm FeH}$ are used. In the absence
of any other data, since ${\rm TiH}$ shares many characteristics
with FeH, this is probably a reasonable assumption. A truncated
Lorentzian profile is used, where the absorption in the wings of
each line is truncated at min(25$\times P_{tot}$,100) cm$^{-1}$ on
either side of the line center, $P_{tot}$ is the total gas pressure
in atmospheres, and the detuning at which the absorption is set to
zero does not exceed 100 cm$^{-1}$. To conserve the total line
strength, the absorption in the part of the profile calculated is
increased to account for the truncated wings. The reason for
truncating the profiles is that we have no satisfactory theory to
deal with the far wings of molecular lines, but we know that for
atomic lines there is a rapid fall-off in the far wings, and, hence
surmise that a rapid fall-off is probably realistic. However, our
choice of determining its location is rather arbitrary.

\begin{figure}
\plotone{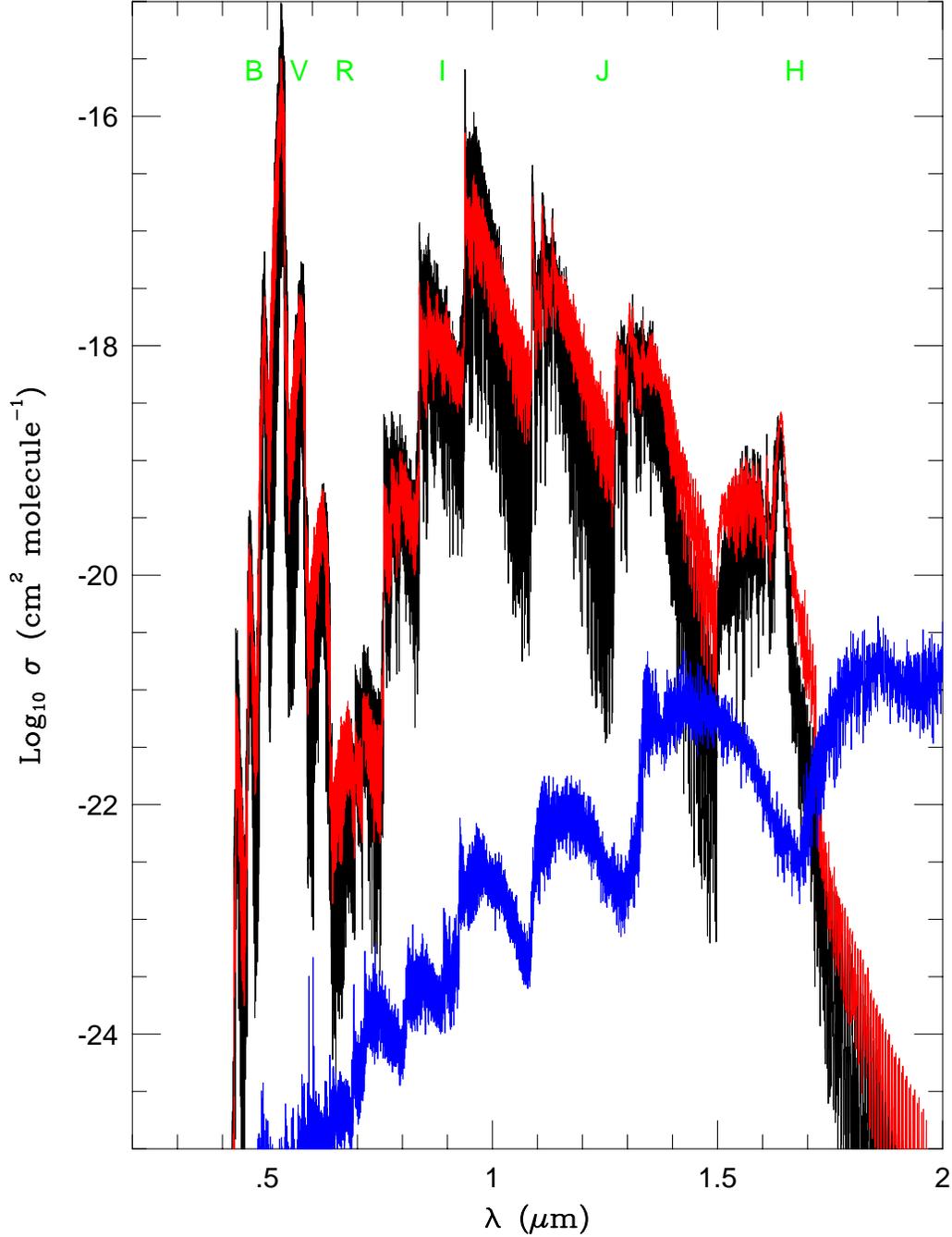} \vspace*{-1.0in}
\caption{\label{fig:t4} The log (base 10) of $\sigma$, the absorption cross section per molecule in cm$^2$,
for ${\rm TiH}$ at 10 atmospheres and 3000 K (red)
and 1 atmosphere and 2000 K (black). For comparison,
the cross section per molecule for ${\rm H_2O}$
at 10 atmospheres and 3000 K is shown in
blue. The approximate positions of the photometric
bands $B$, $V$, $R$, $I$, $J$, and $H$ are
also indicated. The broad absorption of ${\rm TiH}$ with several peaks
between $\sim$0.7 \mic and $\sim$1.95 \mic is due to the $A-X$ electronic band system,
and the absorption between $\sim$0.43 \mic and $\sim$0.7 \mic is due to
the $B-X$ electronic band system.
See text for a detailed discussion.}
\end{figure}

As with FeH, different isotopic versions are included in the
calculations. The main isotope of ${\rm Ti}$ is ${\rm ^{48}Ti}$,
which makes up 73.8\% of terrestrial titanium. The other isotopes of
titanium are ${\rm ^{46}Ti}$, ${\rm ^{47}Ti}$, ${\rm ^{49}Ti}$, and
${\rm ^{50}Ti}$ with isotopic abundances of 8\%, 7.3\%, 5.5\%, and
5.4\%, respectively. Using the same methods as Dulick et al. (2003)
based on Herzberg (1950), the shift in the energy levels, and,
hence, the transition frequencies for the minor isotopic versions of
TiH are calculated. As in the case of Dulick et al., the change in
the line strength with isotope depends only on the isotopic fraction
of Ti in that particular isotopic form, and effects due to changes
in the partition function, Boltzmann factor, and intrinsic line
strength are small, and so are neglected. Figure \ref{fig:t4} shows
the absorption opacity in cm$^2$ per ${\rm TiH}$ molecule (plotted
in red) between a wavelength of 0.3 \mic and 2.0 \mic, at
temperature/pressure points of 3000 K/10 atmospheres and 2000 K/1
atmosphere. For comparison, the ${\rm H_2O}$ (blue) opacities per
molecule at 3000K and 10 atmospheres are also shown. The approximate
location of the photometric bands $B$, $V$, and $R$ in the visible,
and $I$, $J$, and $H$ in the infrared are also indicated.

The absorption spectrum of ${\rm TiH}$ covering the region in the
near infrared from $\sim$0.7 \mic (the limit of the far visible red)
to $\sim$1.95 \mic, is due to the $A-X$ electronic system, and the
absorption between $\sim$0.43 \mic and $\sim$0.7 \mic in the visible
is due to the $B-X$ electronic system. The two electronic systems
overlap at $\sim$0.7 \mic. The strongest absorption in the $A-X$
system corresponds to the $\Delta v$=0 bands (0-0, 1-1, 2-2, etc.),
the next strongest peaks at a shorter wavelength and corresponds to
the $\Delta v=1$ bands (1-0, 2-1, 3-2, etc.), and the next strongest
peaks at a longer wavelength and corresponds to the $\Delta v=-1$
bands (0-1, 1-2, 2-3, etc.).  There is a similar structure for the
$B-X$ system. Many of the peak absorption features of TiH occur at
about the same wavelengths as the peaks in absorption of ${\rm
TiO}$, which is more abundant. This is particularly obvious for the
$A-X$ system of ${\rm TiH}$. Moreover, although the cross section
per molecule of ${\rm H_2O}$ is much less than that of ${\rm TiH}$
over most of the wavelength region, ${\rm H_2O}$ is far more
abundant, and the five peaks corresponding to the $\Delta v$=-2, -1,
0, 1 and 2 sequence of bands for this $A-X$ system of TiH match up
approximately with peaks in the ${\rm H_2O}$ absorption spectrum.

Therefore, because of the particular wavelength positions of the
${\rm TiO}$ and ${\rm H_2O}$ absorption bands and the high
abundances of the ${\rm TiO}$ and ${\rm H_2O}$ molecules at solar
metallicity, identifying ${\rm TiH}$ in a solar-metallicity M or L
dwarf spectrum, while not impossible, is likely to be difficult.
However, the TiH features are not completely obscured, in particular
at $\sim$0.52 \mic, $\sim$0.94 \mic, and weakly in the $H$ band near
1.6 \mic. Notice that the band heads of the $A-X$ transition are
mainly R-heads because $B^\prime < B^{\prime\prime}$ but for the
$\Delta v=-3$ sequence in the $H$ band the shading changes and
P-heads form. The peak in the ${\rm TiH}$ absorption in the $H$ band
corresponds with a deep trough of the ${\rm TiO}$ absorption, and is
close to a minimum in the ${\rm H_2O}$ absorption. Therefore, in
particular in subdwarfs with sub-solar metallicities, for which the
abundances of metal hydrides come into their own vis \`a vis metal
oxides (Fig. \ref{fig:t3}), spectral features of TiH in the $H$ band
and at 0.94 \mic should emerge. The larger TiH abundances we derive
in \S\ref{EOS} also suggest this. The recent detection in the
subdwarf L dwarf 2MASS J0532 by Burgasser, Kirkpatrick \& L\'{e}pine
(2004) of the TiH feature at 0.94 \mic is in keeping with these
expectations.

\section{CONCLUSIONS}
\label{conclusions}

For the TiH molecule, we have combined ab initio calculations
with spectroscopic measurements to derive new thermochemical
data, new spectral line lists and oscillator strengths,
its abundance (for a given composition, temperature, and
pressure), and its per-molecule absorption opacities.  We find that
with the new partition functions, the abundance of TiH in M and L dwarf atmospheres
is much higher than previously thought, that the TiH/TiO ratio increases strongly
with decreasing metallicity, and that at high enough temperatures the TiH/TiO ratio
can exceed unity.  Furthermore, we conclude that, particularly for subdwarf L and M dwarfs,
spectral features of TiH near 0.94 \mic and in the $H$ band may not be weak.  The recent
putative detection of the 0.94 \mic feature in the L
subdwarf 2MASS J0532 is encouraging in this regard and suggests
that the detection of TiH in other dwarfs may shed
light on their atmospheric titanium chemistry.

\acknowledgements

This work was supported in part by NASA under grants
NAG5-10760 and NNG04GL22G. Support
was also provided by the NASA Laboratory Physics Program and the
Natural Sciences and Engineering Research Council of Canada.
The authors would like to thank R. Freedman for providing
guidance with line broadening parameters.
Furthermore, AB acknowledges support through the Cooperative Agreement \#{NNA04CC07A}
between the University of Arizona/NOAO LAPLACE node and NASA's
Astrobiology Institute.  The line lists themselves are available
in electronic form at http://bernath.uwaterloo.ca/TiH .

\appendix

\section{FREE ENERGY OF A GAS-PHASE SPECIES}
\label{appendix1}

The Gibbs free energy of formation of any gas-phase species, relative to its
constituent element or elements in their monatomic gaseous state, can be
calculated if its spectroscopic data, and those of its constituent elements,
are known. We start with the equilibrium
$$ Y^q + q{\rm e^-} \rightleftharpoons \sum_{i=1}^j n_iX_i \, , $$
where $Y$ represents an atomic or molecular species, and has the electric
charge $q$, which is zero for a neutral species, $X_i$ represents the $i$-th
element in $Y$, $n_i$ is the corresponding stoichiometric coefficient, $j$ is
the number of different elements in $Y$, and ${\rm e^-}$ is the electron.

Based on a generalization of the Saha equation and the law of mass action,
the following equation can be derived to obtain the Gibbs energy of
formation of $Y^q$:
\begin{eqnarray}
\Delta G(Y^q) = -RT \Biggl[\ln\left(Q[Y^q]\right) -
\sum_{i=1}^j n_i \ln\left(Q[X_i]\right)+ \frac{hcE}{kT}
+ \frac{5}{2}(1 + q - N) \ln T \nonumber \\
+ \frac{3}{2}\Biggl\{\ln\left(\sum_{i=1}^j n_i m_i\right) - qm_e\Biggr\}
+ q\left(\ln 2 + \frac{3}{2} \ln m_e \right) \nonumber \\
+ (1 + q + N)\biggl\{\frac{5}{2}\ln k
+ \frac{3}{2}\ln\left[\frac{2\pi}{N_Ah^2}\right] - \ln A_o\biggr\}\Biggr] \, ,
\label{DATA}
\end{eqnarray}
where $Q[Y^q]$ and $Q[X_i]$ are the partition functions corresponding to
$Y^q$ and the atoms $X_i$, respectively, and $E$ is the total energy
(in wavenumbers) to fully dissociate a neutral molecule into its constituent atoms,
if the species is a molecule. For a positive ion, the ionization potential (or,
for a multiply charged positive ion, the sum of the ionization potentials) must
be included in $E$. For a negative ion, the electron affinity must be subtracted
from $E$. In addition, the total number of atoms in the molecule is $N$, and is
given by $\sum_{i=1}^{j}n_i$, i.e., the sum of the number of atoms of each element
over the number of elements. The mass of the electron and the mass of the $i$-th
element are $m_e$ and $m_i$ in amu, respectively, $R$, $N_A$, and $A_o$ are the
gas constant, Avogadro's number, and the standard atmosphere, respectively, and
the other symbols have their usual meanings.

To be rigorously correct, for ions the molecular weight is corrected for
the gain/loss in mass due to the addition/removal of electrons.  For
neutral monatomic species all terms on the right hand side of eq. (\ref{DATA}) cancel yielding
zero, and the energy of formation for free electrons is zero by definition.
For convenience, all the temperature-independent terms (after multiplying
by $-RT$) have been collected together, and depend only on the mass, the
number of atoms, and charge of the species.

In order to incorporate the free energies into our existing database in
a uniform way, eq. (\ref{DATA}) is calculated at 100 K intervals over the
temperature range we are likely to use, then a polynomial fit of the form
\begin{equation}
\Delta G(T) = a/T + b + cT + dT^2 + eT^3
\label{GT}
\end{equation}
is made. The lowest number of coefficients that make the best fit over the
calculated range is adopted. The results for TiH using this procedure
are $a$=-3.046$\times 10^5$, $b$=-4.915$\times 10^4$, $c$=2.325$\times 10^1$, $d$=-1.568$\times 10^{-4}$,
and $e$=4.539$\times 10^{-8}$ for $G(T)$ in calories per mole.

In the case of diatomic neutral molecules, such as ${\rm TiH}$, eq. (\ref{DATA})
simplifies by putting $q=0$, $N=2$, $j=2$ and $n_1=n_2=1$, and the appropriate
partition functions for the molecule and the atoms are used.
In addition, the energy $E$ to fully dissociate the molecule is just the
dissociation potential from the ground vibrational level of the ground
electronic state, which for ${\rm TiH}$ is 2.08 eV, or
16776 cm$^{-1}$.

\clearpage
\begin{table}
\caption{\label{theory1}     Summary of IC-MRCI+Q Spectroscopic Constants.}
\begin{center}
\begin{tabular}{lrrrrr}
\tableline\tableline
\noalign{\vskip 5pt}
State & $r_0$(\AA) & $\omega_e$(cm$^{-1}$) & $\omega_e x_e$(cm$^{-1}$)
& Dipole(debye)$^a$& $T_{00}$(cm$^{-1}$)\\\multispan 6 \hfil valence correlation \hfil \\
\noalign{\vskip 5pt}
\multispan 6 \hfil $A{}^4\Phi-X{}^4\Phi$ calculation \hfil \\
$X{}^4\Phi$&    1.818& 1533.4&    21.94& 2.30& 0\\
$A{}^4\Phi$&  1.896& 1375.9&    23.17& 2.81&  10745\\
\multispan 6 \hfil $B{}^4\Gamma-X{}^4\Phi$ calculation \hfil \\
$X{}^4\Phi$&   1.821& 1527.0&    21.85& 2.13&  0\\
$^4\Pi$&      1.838& 1489.3&    22.31& 1.91&  1694\\
$^4\Delta$&   1.905& 1402.4&    21.82& 1.17&  4082\\
$B{}^4\Gamma$&   1.804& 1577.8&    27.52& 3.19&  17735\\
\noalign{\vskip 5pt}
\multispan 6 \hfil core+valence correlation \hfil \\
\noalign{\vskip 5pt}
\multispan 6 \hfil $A{}^4\Phi-X{}^4\Phi$ calculation \hfil \\
$X{}^4\Phi$&   1.788& 1548.9&    20.23& 2.46& 0\\
$A{}^4\Phi$& 1.888& 1342.6&    21.26& 3.25&    11237\\
\multispan 6 \hfil $B{}^4\Gamma-X{}^4\Phi$ calculation \hfil \\
$X{}^4\Phi$&  1.794& 1529.2&    20.04& 1.52\rlap{$^b$}&  0\\
$^4\Pi$&      1.813& 1487.4&    20.36& 1.40&  1732\\
$^4\Delta$&   1.885& 1398.7&    20.66& 1.24&  4231\\
$B{}^4\Gamma$&   1.764& 1592.5&    27.07& 2.59&  18874\\
\noalign{\vskip 5pt}
\multispan 6 \hfil Experiment \hfil \\
$X{}^4\Phi$&   1.779\rlap{$^c$}& 1385.3\rlap{$^d$}&    & 2.45\rlap{5$^e$}& 0\\
$A{}^4\Phi$& 1.867\rlap{$^c$}& &    &    &    10595\rlap{$^c$}\\
$B{}^4\Gamma$&   1.724\rlap{8$^f$}&  &   & 2.99\rlap{8$^e$}&  18878\rlap{$^f$}\\
\tableline
\end{tabular}
\end{center}
\noindent $^a$ Computed as an expectation value at the IC-MRCI
level.\par \noindent $^b$ Using a finite field approach, the
IC-MRCI(+Q) level yields 2.06(2.17)debye.\par \noindent $^c$
Andersson et al. (2003). Note Launila \& Lindgren (1996) give
1.7847~\AA\ for $r_0$ of the $X$ state.\par \noindent $^d$ Chertihin
\&\ Andrews (1994).\par \noindent $^e$ Steimle et al. (1991).\par
\noindent $^f$ Launila \&\ Lindgren (1996).\par
\end{table}
\clearpage
\begin{deluxetable}{cccc}
\tablenum{2}
\tablewidth{14cm}
\tablecaption{Einstein A$_{v^{\prime} v^{\prime\prime}}$  Values for the Bands of
the $B{}^4\Gamma-X{}^4\Phi$ and $A{}^4\Phi-X{}^4\Phi$ Transitions.}
\tablehead{
\colhead{v$^{\prime}$}& \colhead{v$^{\prime\prime}$}  & \colhead{A (s$^{-1}$) [$B{}^4\Gamma-X{}^4\Phi$]} & \colhead{A (s$^{-1}$) [$A{}^4\Phi-X{}^4\Phi$}]}
\startdata
           0  &   0 &     0.4003$\times 10^{8}$ &      0.2336$\times 10^{7}$\\
           0  &  1  &     0.6103$\times 10^{6}$ &     0.7567$\times 10^{6}$\\
           0  &  2  &     0.1279$\times 10^{5}$ &     0.1776$\times 10^{6}$\\
           0  &  3  &     0.2072$\times 10^{3}$ &     0.3919$\times 10^{5}$\\
           0  &  4  &     0.7146 &      -\\
           0  &  5  &     0.2469 &      -\\
           1  &  0  &     0.3239$\times 10^{6}$ &     0.2577$\times 10^{6}$\\
           1  &  1  &     0.3831$\times 10^{8}$ &     0.1329$\times 10^{7}$\\
           1  &  2  &     0.1105$\times 10^{7}$ &     0.1071$\times 10^{7}$\\
           1  &  3  &     0.3751$\times 10^{5}$ &     0.4074$\times 10^{6}$\\
           1  &  4  &     0.9085$\times 10^{3}$ &      -\\
           1  &  5  &     0.6571$\times 10^{1}$ &      -\\
           2  &  0  &     0.1205$\times 10^{4}$ &     0.8251$\times 10^{4}$\\
           2  &  1  &     0.5105$\times 10^{6}$ &     0.4106$\times 10^{6}$\\
           2  &  2  &     0.3664$\times 10^{8}$ &     0.6651$\times 10^{6}$\\
           2  &  3  &     0.1463$\times 10^{7}$ &     0.1090$\times 10^{7}$\\
           2  &  4  &     0.7186$\times 10^{5}$ &      -\\
           2  &  5  &     0.2383$\times 10^{4}$ &      -\\
           3  &  0  &     0.2718$\times 10^{3}$ &     0.9007$\times 10^{2}$\\
           3  &  1  &     0.8623$\times 10^{4}$ &     0.2273$\times 10^{5}$\\
           3  &  2  &     0.5489$\times 10^{6}$ &     0.4793$\times 10^{6}$\\
           3  &  3  &     0.3499$\times 10^{8}$ &     0.4793$\times 10^{6}$\\
           3  &  4  &     0.1664$\times 10^{7}$ &      -\\
           3  &  5  &     0.1119$\times 10^{6}$ &      -\\
           4  &  0  &      -         &     0.2572\\
           4  &  1  &      -         &     0.3647$\times 10^{3}$\\
           4  &  2  &      -         &     0.4142$\times 10^{5}$\\
           4  &  3  &      -         &     0.4816$\times 10^{6}$\\
\enddata
\label{table:X22}
\end{deluxetable}
\clearpage
\begin{deluxetable}{rcl}
\tablenum{3}
\tabletypesize{\footnotesize}
\tablewidth{0pt}
\tablecaption{$^4\Phi$ Hund's Case (a) Hamiltonian Matrix}
\tablehead{}
\startdata
\multicolumn{3}{c}{$x = J + 1/2$, $t = \sqrt{3(x^2 - 4)}$,
$u = \sqrt{4(x^2 - 9)}$, $v = \sqrt{3(x^2 - 16)}$}\\[5pt]
\multicolumn{3}{c}{$\displaystyle{
\begin{array}{r@{\hspace{5pt}}c@{\hspace{5pt}}l}
T(9/2) &=& T + 9 A / 2 + 2 \lambda + 3 \gamma\\[5pt]
T(7/2) &=& T + 3 A / 2 - 2 \lambda - 2 \gamma\\[5pt]
T(5/2) &=& T - 3 A / 2 - 2 \lambda - 5 \gamma\\[5pt]
T(3/2) &=& T - 9 A / 2 + 2 \lambda - 6 \gamma\\[5pt]
\end{array} }$}\\[5pt]
$<9/2|H|9/2>$ &=& $T(9/2) + (x^2 - 10) B - (x^4 - 17 x^2 + 52) D
+ (x^6 - 21 x^4 + 96 x^2 - 40) H$\\[5pt]
$<9/2|H|7/2>$ &=& $<7/2|H|9/2> \:=\: -v\:[B - \gamma/2 - 2 (x^2 - 5) D
+ (3 x^4 - 23 x^2 + 16) H]$\\[5pt]
$<9/2|H|5/2>$ &=& $<5/2|H|9/2> \:=\: -u v\: [D - (3 x^2 - 4) H]$\\[5pt]
$<9/2|H|3/2>$ &=& $<3/2|H|9/2> \:=\: -t u v H$\\[5pt]
$<7/2|H|7/2>$ &=& $T(7/2) + x^2 B - (x^4 + 7 x^2 - 84) D
+ (x^6 + 21 x^4 - 258 x^2 + 264) H$\\[5pt]
$<7/2|H|5/2>$ &=& $<5/2|H|7/2> \:=\: -u\:[B - \gamma/2 - 2 (x^2 + 3) D
+ (3 x^4 + 28 x^2 - 60) H]$\\[5pt]
$<7/2|H|3/2>$ &=& $<3/2|H|7/2> \:=\: -t u\:(D - (3 x^2 + 14) H)$\\[5pt]
$<5/2|H|5/2>$ &=& $T(5/2) + (x^2 + 6) B - (x^4 + 19 x^2 - 12) D
+ (x^6 + 39 x^4 + 72 x^2 - 456) H$\\[5pt]
$<5/2|H|3/2>$ &=& $<3/2|H|5/2> \:=\: -t\:[B - \gamma/2 - 2 (x^2 + 7) D
+ (3 x^4 + 49 x^2 + 100) H]$\\[5pt]
$<3/2|H|3/2>$ &=& $T(3/2) + (x^2 + 8) B - (x^4 + 19 x^2 + 52) D
+ (x^6 + 33 x^4 + 222 x^2 + 248) H$\\[5pt]
\enddata
\end{deluxetable}
\clearpage
\begin{deluxetable}{rcl}
\tablenum{4}
\tabletypesize{\footnotesize}
\tablewidth{0pt}
\tablecaption{$^4\Gamma$ Hund's Case (a) Hamiltonian Matrix}
\tablehead{}
\startdata
\multicolumn{3}{c}{$x = J + 1/2$, $t = \sqrt{3(x^2 - 25)}$,
$u = \sqrt{4(x^2 - 16)}$, $v = \sqrt{3(x^2 - 9)}$}\\[5pt]
\multicolumn{3}{c}{$\displaystyle{
\begin{array}{r@{\hspace{5pt}}c@{\hspace{5pt}}l}
T(11/2) &=& T + 6 A + 2 \lambda + 9 \gamma / 2\\[5pt]
T(9/2)  &=& T + 2 A - 2 \lambda - 3 \gamma / 2\\[5pt]
T(7/2)  &=& T - 2 A - 2 \lambda - 11 \gamma / 2\\[5pt]
T(5/2)  &=& T - 6 A + 2 \lambda - 15 \gamma / 2
\end{array} }$}\\[5pt]
$<11/2|H|11/2>$ &=& $T(11/2) + (x^2 - 13) B - (x^4 - 23 x^2 + 94) D
+ (x^6 - 30 x^4 + 201 x^2 - 172) H$\\
$<11/2|H|9/2>$ &=& $<9/2|H|11/2> \:=\: -t\:[B - \gamma/2 - 2 (x^2 - 7) D
+ (3 x^4 - 35 x^2 + 44) H]$\\
$<11/2|H|7/2>$ &=& $<7/2|H|11/2> \:=\: -t u\: [D - (3 x^2 - 7) H]$\\
$<11/2|H|5/2>$ &=& $<5/2|H|11/2> \:=\: -t u v H$\\
$<9/2|H|9/2>$ &=& $T(9/2) + (x^2 - 1) B - (x^4 + 5 x^2 - 138) D
+ (x^6 + 18 x^4 - 439 x^2 + 804) H$\\
$<9/2|H|7/2>$ &=& $<7/2|H|9/2> \:=\: -u \:[B - \gamma/2 - 2 (x^2 + 3) D
+ (3 x^4 + 28 x^2 - 123) H]$\\
$<9/2|H|5/2>$ &=& $<5/2|H|9/2> \:=\: -u v \:[D - (3 x^2 + 17) H]$\\
$<7/2|H|7/2>$ &=& $T(7/2) + (x^2 + 7) B - (x^4 + 21 x^2 - 42) D
+ (x^6 + 42 x^4 + x^2 - 1164) H$\\
$<7/2|H|5/2>$ &=& $<5/2|H|7/2> \:=\: -v\:[B - \gamma/2 - 2 (x^2 + 9) D
+ (3 x^4 + 61 x^2 + 156) H]$\\
$<5/2|H|5/2>$ &=& $T(5/2) + (x^2 + 11) B - (x^4 + 25 x^2 + 94) D
+ (x^6 + 42 x^4 + 369 x^2 + 548) H$\\
\enddata
\end{deluxetable}
\clearpage
\begin{deluxetable}{cllllll}
\tabletypesize{\footnotesize}
\tablenum{5}
\tablewidth{0pt}
\tablecaption{Derived Molecular Constants (cm$^{-1}$) From
Observed Data}
\tablehead{\colhead{State}
& \colhead{$T_0$}
& \colhead{$B_0$}
& \colhead{$D_0\times10^4$}
& \colhead{$H_0\times10^9$}
& \colhead{$A_0$}
& \colhead{$\gamma_0$}}
\startdata
$X^4\Phi$ & \phm{106}85.1362 & 5.3668672(614) & 2.61187(303)
& \phm{--}9.108(405) &
32.96454(797) & \phm{--}0.193432(889)\\
$A^4\Phi(e)$\tablenotemark{a} & 10688.3028(848) & 4.876910(846) & 2.58010 &
--8.025 &
30.4589(246) & --0.10251(733)\\
$A^4\Phi(f)$\tablenotemark{a} & 10687.880(156) & 4.86282(109) & 2.55780 &
--8.112 &
29.0992(410) & --0.22281(435)\\
$B^4\Gamma$ & 18799.595(238) & 5.70361(115) & 2.94312 & --7.459 &
39.1548(442) & --1.0047(147)\\
\enddata
\tablenotetext{a}{Spin-spin constants, $\lambda(e)=-1.599(109)$
and $\lambda(f)=-5.586(118)$.}
\end{deluxetable}
\clearpage
\begin{deluxetable}{cccccccc}
\tablenum{6}
\tabletypesize{\footnotesize}
\tablewidth{0pt}
\tablecaption{Generated Molecular Constants (cm$^{-1}$) for
the $X\:^4\Phi$, $A\:^4\Phi$, and $B\:^4\Gamma$ Vibrational Levels}
\tablehead{\colhead{$v$} & \colhead{$B_v$} & \colhead{$D_v\times10^4$}
& \colhead{$H_v\times10^9$} & \colhead{$T_v(\Lambda-3/2)$}
& \colhead{$T_v(\Lambda-1/2)$} & \colhead{$T_v(\Lambda+1/2)$}
& \colhead{$T_v(\Lambda+3/2)$}}
\startdata
\multicolumn{8}{c}{$X\:^4\Phi$ State\tablenotemark{a}}\\
0 & 5.3668672 & 2.61187 &\phm{--}9.108 &\ --64.3649 &\phm{15}34.7222
&\phm{1}134.1962 & \phm{1}234.0570\\
1 & 5.1007701 & 2.21602 & --8.378 & 1444.1222 & 1543.2093 & 1642.6833 &
1742.5441\\
2 & 4.8611877 & 1.91820 & --8.781 & 2912.1487 & 3011.2358 & 3110.7098 &
3210.5707\\
3 & 4.6513671 & 1.68039 & --8.813 & 4338.4489 & 4437.5360 & 4537.0100 &
4636.8709\\
4 & 4.4745555 & 1.49595 & --8.652 & 5721.7571 & 5820.8442 & 5920.3182 &
6020.1790\\
5 & 4.3340001 & 1.35936 & --8.424 & 7060.8075 & 7159.8946 & 7259.3686 &
7359.2294\\
\multicolumn{8}{c}{$A\:^4\Phi(e)$ State\tablenotemark{b}}\\
0 & 4.8769102 & 2.58010 & --8.025 & 10548.6548 & 10646.3250 &
10737.3941 & 10821.8623\\
1 & 4.6169393 & 2.18919 & --9.213 & 11848.7691 & 11946.4393 &
12037.5085 & 12121.9766\\
2 & 4.3736732 & 1.86106 & --9.643 & 13106.3657 & 13204.0358 &
13295.1050 & 13379.5732\\
3 & 4.1484432 & 1.58810 & --9.585 & 14319.7104 & 14417.3805 &
14508.4497 & 14592.9179\\
4 & 3.9425804 & 1.36321 & --9.237 & 15487.0692 & 15584.7394 &
15675.8086 & 15760.2767\\
5 & 3.7574161 & 1.18002 & --8.745 & 16606.7082 & 16704.3784 &
16795.4476 & 16879.9157\\
\multicolumn{8}{c}{$A\:^4\Phi(f)$ State\tablenotemark{b}}\\
0 & 4.8628162 & 2.55780 & --8.112 & 10547.0989 & 10656.5177 &
10743.1469 & 10806.9864\\
1 & 4.6028453 & 2.16921 & --9.254 & 11847.2133 & 11956.6321 &
12043.2612 & 12107.1008\\
2 & 4.3595792 & 1.84313 & --9.651 & 13104.8098 & 13214.2286 &
13300.8578 & 13364.6973\\
3 & 4.1343492 & 1.57197 & --9.569 & 14318.1545 & 14427.5733 &
14514.2025 & 14578.0420\\
4 & 3.9284864 & 1.34864 & --9.205 & 15485.5134 & 15594.9321 &
15681.5613 & 15745.4009\\
5 & 3.7433221 & 1.16679 & --8.701 & 16605.1524 & 16714.5711 &
16801.2003 & 16865.0399\\
\\\\\\\\\\
\multicolumn{8}{c}{$B\:^4\Gamma$ State\tablenotemark{c}}\\
0 & 5.7036054 & 2.94312 & --7.459 & 18572.2017 & 18726.8115 &
18879.4119 & 19030.0029\\
1 & 5.4171527 & 2.52009 & --8.938 & 20110.5248 & 20265.1346 &
20417.7350 & 20568.3260\\
2 & 5.1537792 & 2.17010 & --9.613 & 21594.7116 & 21749.3214 &
21901.9218 & 22052.5128\\
3 & 4.9169618 & 1.88449 & --9.772 & 23020.5977 & 23175.2075 &
23327.8079 & 23478.3989\\
4 & 4.7101772 & 1.65659 & --9.636 & 24384.0185 & 24538.6283 &
24691.2287 & 24841.8197\\
5 & 4.5369023 & 1.48041 & --9.363 & 25680.8094 & 25835.4192 &
25988.0196 & 26138.6106\\
\enddata
\tablenotetext{a}{$\omega_e = 1547.7347$, $\omega_e x_e = 19.2810$,
$\omega_e y_e = -0.210954$, $B_e = 5.5088441$, $\alpha_e = 0.2895$,
$\gamma_e = 0.010822$, $\epsilon_e = 0.00054119$, and
$\gamma_{0-5} = 0.193432$.}
\tablenotetext{b}{$\omega_e = 1340.9704$, $\omega_e x_e = 19.9584$,
$\omega_e y_e = -0.289002$, $B_e(e) = 5.012744$, $B_e(f) = 4.99865$,
$\alpha_e = 0.2754$, $\gamma_e = 0.007354$, $\epsilon_e = 0.00022187$,
$\gamma_{0-5}(e) = -0.10251$, and $\gamma_{0-5}(f) = -0.22281$.}
\tablenotetext{c}{$\omega_e = 1588.4683$, $\omega_e x_e = 23.9447$,
$\omega_e y_e = -0.694092$, $B_e = 5.8544$, $\alpha_e = 0.3062$,
$\gamma_e = 0.008932$, $\epsilon_e = 0.00057947$, $\gamma_{0-5} = -1.0047$.}
\end{deluxetable}
\clearpage
\begin{deluxetable}{cccccl}
\tablenum{7}
\tabletypesize{\footnotesize}
\tablewidth{0pt}
\tablecaption{${^4\Phi} - {^4\Phi}$ H\"{o}nl-London Factors -- Case (a)
Coupling}
\tablehead{\colhead{Branch} & \colhead{$\Omega^{\prime\prime}$}
& \colhead{Parity} & \colhead{$\Omega^{\prime}$} & \colhead{Parity}
& \colhead{$S(J)$}}
\startdata
$ P(J) $ & 9/2 &$e/f$& 9/2 &$e/f$& $\displaystyle{\frac{1}{4}\cdot
\frac{(2 J - 9) \: (2 J + 9)}{J}}$\\[10pt]
$ P(J) $ & 7/2 &$e/f$& 7/2 &$e/f$& $\displaystyle{\frac{1}{4}\cdot
\frac{(2 J - 7) \: (2 J + 7)}{J}}$\\[10pt]
$ P(J) $ & 5/2 &$e/f$& 5/2 &$e/f$& $\displaystyle{\frac{1}{4}\cdot
\frac{(2 J - 5) \: (2 J + 5)}{J}}$\\[10pt]
$ P(J) $ & 3/2 &$e/f$& 3/2 &$e/f$& $\displaystyle{\frac{1}{4}\cdot
\frac{(2 J - 3) \: (2 J + 3)}{J}}$\\[10pt]
$ Q(J) $ & 9/2 &$e/f$& 9/2 &$f/e$& $\displaystyle{\frac{81}{4}\cdot
\frac{2 J + 1}{J \: (J + 1)}}$\\[10pt]
$ Q(J) $ & 7/2 &$e/f$& 7/2 &$f/e$& $\displaystyle{\frac{49}{4}\cdot
\frac{2 J + 1}{J \: (J + 1)}}$\\[10pt]
$ Q(J) $ & 5/2 &$e/f$& 5/2 &$f/e$& $\displaystyle{\frac{25}{4}\cdot
\frac{2 J + 1}{J \: (J + 1)}}$\\[10pt]
$ Q(J) $ & 3/2 &$e/f$& 3/2 &$f/e$& $\displaystyle{\frac{9}{4}\cdot
\frac{2 J + 1}{J \: (J + 1)}}$\\[10pt]
$ R(J) $ & 9/2 &$e/f$& 9/2 &$e/f$& $\displaystyle{\frac{1}{4}\cdot
\frac{(2 J - 7) \: (2 J + 11)}{J + 1}}$\\[10pt]
$ R(J) $ & 7/2 &$e/f$& 7/2 &$e/f$& $\displaystyle{\frac{1}{4}\cdot
\frac{(2 J - 5) \: (2 J + 9)}{J + 1}}$\\[10pt]
$ R(J) $ & 5/2 &$e/f$& 5/2 &$e/f$& $\displaystyle{\frac{1}{4}\cdot
\frac{(2 J - 3) \: (2 J + 7)}{J + 1}}$\\[10pt]
$ R(J) $ & 3/2 &$e/f$& 3/2 &$e/f$& $\displaystyle{\frac{1}{4}\:
\frac{(2 J - 1) \: (2 J + 5)}{J + 1}}$\\[10pt]
\enddata
\end{deluxetable}
\clearpage
\begin{deluxetable}{cccccl}
\tablenum{8}
\tabletypesize{\footnotesize}
\tablewidth{438pt}
\tablecaption{${^4\Phi} - {^4\Phi}$ H\"{o}nl-London Factors -- Case (b)
Coupling}
\tablehead{\colhead{Branch} & \colhead{$N^{\prime\prime}$}
& \colhead{Parity} & \colhead{$N^{\prime}$} & \colhead{Parity}
& \colhead{$S(J)$}}
\startdata
$ ^PP(J) $ & $J - 3/2$ & $+/-$ & $J - 5/2$ & $+/-$
& $\displaystyle{\frac{1}{4}\cdot\frac{(2 J - 9)\:(2 J + 1)\:(2 J + 3)}
{(2 J - 3)\:(J - 1)}}$\\[10pt]
$ ^QP(J) $ & $J - 3/2$ & $+/-$ & $J - 3/2$ & $+/-$
& $\displaystyle{108\cdot\frac{2 J + 1}{J\:( 2 J - 3)\:(2 J - 1)^2}}$\\[10pt]
$ ^RP(J) $ & $J - 3/2$ & $+/-$ & $J - 1/2$ & $+/-$
& $\displaystyle{\frac{3}{4}\cdot\frac{(2 J - 7)\:(2 J + 5)}
{J^2\:(2 J - 1)^2\:(J - 1)}}$\\[10pt]
$ ^PP(J) $ & $J - 1/2$ & $+/-$ & $J - 3/2$ & $+/-$
& $\displaystyle{\frac{1}{4}\cdot
\frac{(2 J - 7)\:(2 J - 3)\:(2 J + 1)\:(2 J + 5)\:(J + 1)}
{J^2\:(2 J - 1)^2}}$\\[10pt]
$ ^QP(J) $ & $J - 1/2$ & $+/-$ & $J - 1/2$ & $+/-$
& $\displaystyle{576\cdot\frac{(J - 1)\:(J + 1)}
{J\:(2 J - 1)^2\:(2 J + 1)^2}}$\\[10pt]
$ ^RP(J) $ & $J - 1/2$ & $+/-$ & $J + 1/2$ & $+/-$
& $\displaystyle{\frac{3}{4}\cdot\frac{(2 J - 5)\:(2 J + 7)}
{J^2\:(2 J + 1)^2\:(J + 1)}}$\\[10pt]
$ ^PP(J) $ & $J + 1/2$ & $+/-$ & $J - 1/2$ & $+/-$
& $\displaystyle{\frac{1}{4}\cdot
\frac{(2 J - 5)\:(2 J - 1)\:(2 J + 3)\:(2 J + 7)\:(J - 1)}
{J^2\:(2 J + 1)^2}}$\\[10pt]
$ ^QP(J) $ & $J + 1/2$ & $+/-$ & $J + 1/2$ & $+/-$
& $\displaystyle{108\cdot\frac{2 J - 1}{J\:(2 J + 1)^2\:(2 J + 3)}}$\\[10pt]
$ ^PP(J) $ & $J + 3/2$ & $+/-$ & $J + 1/2$ & $+/-$
& $\displaystyle{\frac{1}{4}\cdot\frac{(2 J - 3)\:(2 J - 1)\:(2 J + 9)}
{(2 J + 3)\:(J + 1)}}$\\[10pt]
$ ^QQ(J) $ & $J - 3/2$ & $+/-$ & $J - 3/2$ & $+/-$
& $\displaystyle{36\cdot\frac{(2 J + 1)\:(J + 1)}
{J\:(2 J - 1)^2}}$\\[10pt]
$ ^RQ(J) $ & $J - 3/2$ & $+/-$ & $J - 1/2$ & $+/-$
& $\displaystyle{\frac{3}{4}\cdot\frac{(2 J - 7)\:(2 J + 1)\:(2 J + 5)}
{J^2\:(2 J - 1)^2}}$\\[10pt]
$ ^PQ(J) $ & $J - 1/2$ & $+/-$ & $J - 3/2$ & $+/-$
& $\displaystyle{\frac{3}{4}\cdot\frac{(2 J - 7)\:(2 J + 1)\:(2 J + 5)}
{J^2\:(2 J - 1)^2}}$\\[10pt]
$ ^QQ(J) $ & $J - 1/2$ & $+/-$ & $J - 1/2$ & $+/-$
& $\displaystyle{36\cdot\frac{(2 J^2 + J - 4)^2}
{J\:(2 J - 1)^2\:(2 J + 1)\:(J + 1)}}$\\[10pt]
$ ^RQ(J) $ & $J - 1/2$ & $+/-$ & $J + 1/2$ & $+/-$
& $\displaystyle{\frac{1}{4}\cdot
\frac{(2 J - 5)\:(2 J - 1)\:(2 J + 3)\:(2 J + 7)}
{J^2\:(2 J + 1)\:(J + 1)^2}}$\\[10pt]
$ ^PQ(J) $ & $J + 1/2$ & $+/-$ & $J - 1/2$ & $+/-$
& $\displaystyle{\frac{1}{4}\cdot
\frac{(2 J - 5)\:(2 J - 1)\:(2 J + 3)\:(2 J + 7)}
{J^2\:(2 J + 1)\:(J + 1)^2}}$\\[10pt]
$ ^QQ(J) $ & $J + 1/2$ & $+/-$ & $J + 1/2$ & $+/-$
& $\displaystyle{36\cdot\frac{(2 J^2 + 3 J - 3)^2}
{J\:(2 J + 1)\:(2 J + 3)^2\:(J + 1)}}$\\[10pt]
\\\\\\\\\\\\\\\\\\\\
$ ^RQ(J) $ & $J + 1/2$ & $+/-$ & $J + 3/2$ & $+/-$
& $\displaystyle{\frac{3}{4}\cdot
\frac{(2 J - 3)\:(2 J + 1)\:(2 J + 9)}
{(2 J + 3)^2\:(J + 1)^2}}$\\[10pt]
$ ^PQ(J) $ & $J + 3/2$ & $+/-$ & $J + 1/2$ & $+/-$
& $\displaystyle{\frac{3}{4}\cdot
\frac{(2 J - 3)\:(2 J + 1)\:(2 J + 9)}
{(2 J + 3)^2\:(J + 1)^2}}$\\[10pt]
$ ^QQ(J) $ & $J + 3/2$ & $+/-$ & $J + 3/2$ & $+/-$
& $\displaystyle{36\cdot
\frac{J\:(2 J + 1)}
{(2 J + 3)^2\:(J + 1)}}$\\[10pt]
$ ^RR(J) $ & $J - 3/2$ & $+/-$ & $J - 1/2$ & $+/-$
& $\displaystyle{\frac{1}{4}\cdot\frac{(2 J - 7)\:(2 J + 3)\:(2 J + 5)}
{J\:(2 J - 1)}}$\\[10pt]
$ ^QR(J) $ & $J - 1/2$ & $+/-$ & $J - 1/2$ & $+/-$
& $\displaystyle{108\cdot\frac{2 J + 3}{(2 J - 1)\:(2 J + 1)^2\:(J + 1)}}$\\
[10pt]
$ ^RR(J) $ & $J - 1/2$ & $+/-$ & $J + 1/2$ & $+/-$
& $\displaystyle{\frac{1}{4}\cdot
\frac{(2 J - 5)\:(2 J - 1)\:(2 J + 3)\:(2 J + 7)\:(J + 2)}
{(2 J + 1)^2\:(J + 1)^2}}$\\[10pt]
$ ^PR(J) $ & $J + 1/2$ & $+/-$ & $J - 1/2$ & $+/-$
& $\displaystyle{\frac{3}{4}\cdot\frac{(2 J - 5)\:(2 J + 7)}
{J\:(2 J + 1)^2\:(J + 1)^2}}$\\[10pt]
$ ^QR(J) $ & $J + 1/2$ & $+/-$ & $J + 1/2$ & $+/-$
& $\displaystyle{576\cdot\frac{J\:(J + 2)}
{(2 J + 1)^2\:(2 J + 3)^2\:(J + 1)}}$\\[10pt]
$ ^RR(J) $ & $J + 1/2$ & $+/-$ & $J + 3/2$ & $+/-$
& $\displaystyle{\frac{1}{4}\cdot
\frac{J\:(2 J - 3)\:(2 J + 1)\:(2 J + 5)\:(2 J + 9)}
{(2 J + 3)^2\:(J + 1)^2}}$\\[10pt]
$ ^PR(J) $ & $J + 3/2$ & $+/-$ & $J + 1/2$ & $+/-$
& $\displaystyle{\frac{3}{4}\cdot\frac{(2 J - 3)\:(2 J + 9)}
{(2 J + 3)^2\:(J + 1)^2\:(J + 2)}}$\\[10pt]
$ ^QR(J) $ & $J + 3/2$ & $+/-$ & $J + 3/2$ & $+/-$
& $\displaystyle{108\cdot\frac{2 J + 1}
{(J + 1)\:(2 J + 3)^2\:(2 J + 5)}}$\\[10pt]
$ ^RR(J) $ & $J + 3/2$ & $+/-$ & $J + 5/2$ & $+/-$
& $\displaystyle{\frac{1}{4}\cdot\frac{(2 J - 1)\:(2 J + 1)\:(2 J + 11)}
{(2 J + 5)\:(J + 2)}}$\\[10pt]
\enddata
\end{deluxetable}
\clearpage
\begin{deluxetable}{cccccl}
\tablenum{9}
\tabletypesize{\footnotesize}
\tablewidth{0pt}
\tablecaption{${^4\Gamma} - {^4\Phi}$ H\"{o}nl-London
Factors -- Case (a) Coupling}
\tablehead{\colhead{Branch} & \colhead{$\Omega^{\prime\prime}$}
& \colhead{Parity} & \colhead{$\Omega^{\prime}$} & \colhead{Parity}
& \colhead{$S(J)$}}
\startdata
$ P(J) $ & 9/2 &$e/f$& 11/2 &$e/f$& $\displaystyle{\frac{1}{8}\cdot
\frac{(2 J - 11) \: (2 J - 9)}{J}}$\\[10pt]
$ P(J) $ & 7/2 &$e/f$& 9/2 &$e/f$& $\displaystyle{\frac{1}{8}\cdot
\frac{(2 J - 9) \: (2 J - 7)}{J}}$\\[10pt]
$ P(J) $ & 5/2 &$e/f$& 7/2 &$e/f$& $\displaystyle{\frac{1}{8}\cdot
\frac{(2 J - 7) \: (2 J - 5)}{J}}$\\[10pt]
$ P(J) $ & 3/2 &$e/f$& 5/2 &$e/f$& $\displaystyle{\frac{1}{8}\cdot
\frac{(2 J - 5) \: (2 J - 3)}{J}}$\\[10pt]
$ Q(J) $ & 9/2 &$e/f$& 11/2 &$f/e$& $\displaystyle{\frac{1}{8}\cdot
\frac{(2 J - 9)\:(2 J + 1)\:(2 J + 11)}{J \: (J + 1)}}$\\[10pt]
$ Q(J) $ & 7/2 &$e/f$& 9/2 &$f/e$& $\displaystyle{\frac{1}{8}\cdot
\frac{(2 J - 7)\:(2 J + 1)\:(2 J + 9)}{J \: (J + 1)}}$\\[10pt]
$ Q(J) $ & 5/2 &$e/f$& 7/2 &$f/e$& $\displaystyle{\frac{1}{8}\cdot
\frac{(2 J - 5)\:(2 J + 1)\:(2 J + 7)}{J \: (J + 1)}}$\\[10pt]
$ Q(J) $ & 3/2 &$e/f$& 5/2 &$f/e$& $\displaystyle{\frac{1}{8}\cdot
\frac{(2 J - 3)\:(2 J + 1)\:(2 J + 5)}{J \: (J + 1)}}$\\[10pt]
$ R(J) $ & 9/2 &$e/f$& 11/2 &$e/f$& $\displaystyle{\frac{1}{8}\cdot
\frac{(2 J + 11)\:(2 J + 13)}{J + 1}}$\\[10pt]
$ R(J) $ & 7/2 &$e/f$& 9/2 &$e/f$& $\displaystyle{\frac{1}{8}\cdot
\frac{(2 J + 9) \: (2 J + 11)}{J + 1}}$\\[10pt]
$ R(J) $ & 5/2 &$e/f$& 7/2 &$e/f$& $\displaystyle{\frac{1}{8}\cdot
\frac{(2 J + 7) \: (2 J + 9)}{J + 1}}$\\[10pt]
$ R(J) $ & 3/2 &$e/f$& 5/2 &$e/f$& $\displaystyle{\frac{1}{8}\cdot
\frac{(2 J + 5) \: (2 J + 7)}{J + 1}}$\\
\enddata
\end{deluxetable}
\clearpage
\begin{deluxetable}{cccccl}
\tablenum{10}
\tabletypesize{\footnotesize}
\tablewidth{438pt}
\tablecaption{${^4\Gamma} - {^4\Phi}$ H\"{o}nl-London
Factors -- Case (b) Coupling}
\tablehead{\colhead{Branch} & \colhead{$N^{\prime\prime}$}
& \colhead{Parity} & \colhead{$N^{\prime}$} & \colhead{Parity}
& \colhead{$S(J)$}}
\startdata
$ ^PP(J) $ & $J - 3/2$ & $+/-$ & $J - 5/2$ & $+/-$
& $\displaystyle{\frac{1}{8}\cdot\frac{(2 J - 11)\:(2 J - 9)\:(2 J + 1)}
{(2 J - 3)\:(J - 1)}}$\\[10pt]
$ ^QP(J) $ & $J - 3/2$ & $+/-$ & $J - 3/2$ & $+/-$
& $\displaystyle{\frac{3}{2}\cdot\frac{(2 J - 9)\:(2 J + 1)\:(2 J + 5)}
{J\:( 2 J - 3)\:(2 J - 1)^2}}$\\[10pt]
$ ^RP(J) $ & $J - 3/2$ & $+/-$ & $J - 1/2$ & $+/-$
& $\displaystyle{\frac{3}{8}\cdot\frac{(2 J + 5)\:(2 J + 7)}
{J^2\:(2 J - 1)^2\:(J - 1)}}$\\[10pt]
$ ^PP(J) $ & $J - 1/2$ & $+/-$ & $J - 3/2$ & $+/-$
& $\displaystyle{\frac{1}{8}\cdot
\frac{(2 J - 9)\:(2 J - 7)\:(2 J - 3)\:(2 J + 1)\:(J + 1)}
{J^2\:(2 J - 1)^2}}$\\[10pt]
$ ^QP(J) $ & $J - 1/2$ & $+/-$ & $J - 1/2$ & $+/-$
& $\displaystyle{8\cdot\frac{(2 J - 7)\:(2 J + 7)\:(J - 1)\:(J + 1)}
{J\:(2 J - 1)^2\:(2 J + 1)^2}}$\\[10pt]
$ ^RP(J) $ & $J - 1/2$ & $+/-$ & $J + 1/2$ & $+/-$
& $\displaystyle{\frac{3}{8}\cdot\frac{(2 J + 7)\:(2 J + 9)}
{J^2\:(2 J + 1)^2\:(J + 1)}}$\\[10pt]
$ ^PP(J) $ & $J + 1/2$ & $+/-$ & $J - 1/2$ & $+/-$
& $\displaystyle{\frac{1}{8}\cdot
\frac{(2 J - 7)\:(2 J - 5)\:(2 J - 1)\:(2 J + 3)\:(J - 1)}
{J^2\:(2 J + 1)^2}}$\\[10pt]
$ ^QP(J) $ & $J + 1/2$ & $+/-$ & $J + 1/2$ & $+/-$
& $\displaystyle{\frac{3}{2}\cdot\frac{(2 J - 5)\:(2 J - 1)\:(2 J + 9)}
{J\:(2 J + 1)^2\:(2 J + 3)}}$\\[10pt]
$ ^PP(J) $ & $J + 3/2$ & $+/-$ & $J + 1/2$ & $+/-$
& $\displaystyle{\frac{1}{8}\cdot\frac{(2 J - 5)\:(2 J - 3)\:(2 J - 1)}
{(2 J + 3)\:(J + 1)}}$\\[10pt]
$ ^QQ(J) $ & $J - 3/2$ & $+/-$ & $J - 3/2$ & $+/-$
& $\displaystyle{\frac{1}{2}\cdot
\frac{(2 J - 9)\:(2 J + 5)\:(2 J + 1)\:(J + 1)}{J\:(2 J - 1)^2}}$\\[10pt]
$ ^RQ(J) $ & $J - 3/2$ & $+/-$ & $J - 1/2$ & $+/-$
& $\displaystyle{\frac{3}{8}\cdot\frac{(2 J + 1)\:(2 J + 5)\:(2 J + 7)}
{J^2\:(2 J - 1)^2}}$\\[10pt]
$ ^PQ(J) $ & $J - 1/2$ & $+/-$ & $J - 3/2$ & $+/-$
& $\displaystyle{\frac{3}{8}\cdot\frac{(2 J - 9)\:(2 J - 7)\:(2 J + 1)}
{J^2\:(2 J - 1)^2}}$\\[10pt]
$ ^QQ(J) $ & $J - 1/2$ & $+/-$ & $J - 1/2$ & $+/-$
& $\displaystyle{\frac{1}{2}\cdot
\frac{(2 J - 7)\:(2 J^2 + J - 4)^2\:(2 J + 7)}
{J\:(2 J - 1)^2\:(2 J + 1)\:(J + 1)}}$\\[10pt]
$ ^RQ(J) $ & $J - 1/2$ & $+/-$ & $J + 1/2$ & $+/-$
& $\displaystyle{\frac{1}{8}\cdot
\frac{(2 J - 1)\:(2 J + 3)\:(2 J + 7)\:(2 J + 9)}
{J^2\:(2 J + 1)\:(J + 1)^2}}$\\[10pt]
$ ^PQ(J) $ & $J + 1/2$ & $+/-$ & $J - 1/2$ & $+/-$
& $\displaystyle{\frac{1}{8}\cdot
\frac{(2 J - 7)\:(2 J - 5)\:(2 J - 1)\:(2 J + 3)}
{J^2\:(2 J + 1)\:(J + 1)^2}}$\\[10pt]
$ ^QQ(J) $ & $J + 1/2$ & $+/-$ & $J + 1/2$ & $+/-$
& $\displaystyle{\frac{1}{2}\cdot
\frac{(2 J - 5)\:(2 J^2 + 3 J - 3)^2\:(2 J + 9)}
{J\:(2 J + 1)\:(2 J + 3)^2\:(J + 1)}}$\\[10pt]
\\\\\\\\\\\\\\\\\\\\
$ ^RQ(J) $ & $J + 1/2$ & $+/-$ & $J + 3/2$ & $+/-$
& $\displaystyle{\frac{3}{8}\cdot
\frac{(2 J + 1)\:(2 J + 9)\:(2 J + 11)}
{(2 J + 3)^2\:(J + 1)^2}}$\\[10pt]
$ ^PQ(J) $ & $J + 3/2$ & $+/-$ & $J + 1/2$ & $+/-$
& $\displaystyle{\frac{3}{8}\cdot
\frac{(2 J - 5)\:(2 J - 3)\:(2 J + 1)}
{(2 J + 3)^2\:(J + 1)^2}}$\\[10pt]
$ ^QQ(J) $ & $J + 3/2$ & $+/-$ & $J + 3/2$ & $+/-$
& $\displaystyle{\frac{1}{2}\cdot
\frac{J\:(2 J - 3)\:(2 J + 1)\:(2 J + 11)}
{(2 J + 3)^2\:(J + 1)}}$\\[10pt]
$ ^RR(J) $ & $J - 3/2$ & $+/-$ & $J - 1/2$ & $+/-$
& $\displaystyle{\frac{1}{8}\cdot
\frac{(2 J + 3)\:(2 J + 5)\:(2 J + 7)}
{J\:(2 J - 1)}}$\\[10pt]
$ ^QR(J) $ & $J - 1/2$ & $+/-$ & $J - 1/2$ & $+/-$
& $\displaystyle{\frac{3}{2}\cdot
\frac{(2 J - 7)\:(2 J + 3)\:(2 J + 7)}
{(2 J - 1)\:(2 J + 1)^2\:(J + 1)}}$\\[10pt]
$ ^RR(J) $ & $J - 1/2$ & $+/-$ & $J + 1/2$ & $+/-$
& $\displaystyle{\frac{1}{8}\cdot
\frac{(2 J - 1)\:(2 J + 3)\:(2 J + 7)\:(2 J + 9)\:(J + 2)}
{(2 J + 1)^2\:(J + 1)^2}}$\\[10pt]
$ ^PR(J) $ & $J + 1/2$ & $+/-$ & $J - 1/2$ & $+/-$
& $\displaystyle{\frac{3}{8}\cdot
\frac{(2 J - 7)\:(2 J - 5)}
{J\:(2 J + 1)^2\:(J + 1)^2}}$\\[10pt]
$ ^QR(J) $ & $J + 1/2$ & $+/-$ & $J + 1/2$ & $+/-$
& $\displaystyle{8\cdot
\frac{J\:(2 J - 5)\:(2 J + 9)\:(J + 2)}
{(2 J + 1)^2\:(2 J + 3)^2\:(J + 1)}}$\\[10pt]
$ ^RR(J) $ & $J + 1/2$ & $+/-$ & $J + 3/2$ & $+/-$
& $\displaystyle{\frac{1}{8}\cdot
\frac{J\:(2 J + 1)\:(2 J + 5)\:(2 J + 9)\:(2 J + 11)}
{(2 J + 3)^2\:(J + 1)^2}}$\\[10pt]
$ ^PR(J) $ & $J + 3/2$ & $+/-$ & $J + 1/2$ & $+/-$
& $\displaystyle{\frac{3}{8}\cdot
\frac{(2 J - 5)\:(2 J - 3)}
{(2 J + 3)^2\:(J + 1)^2\:(J + 2)}}$\\[10pt]
$ ^QR(J) $ & $J + 3/2$ & $+/-$ & $J + 3/2$ & $+/-$
& $\displaystyle{\frac{3}{2}\cdot
\frac{(2 J - 3)\:(2 J + 1)\:(2 J + 11)}
{(J + 1)\:(2 J + 3)^2\:(2 J + 5)}}$\\[10pt]
$ ^RR(J) $ & $J + 3/2$ & $+/-$ & $J + 5/2$ & $+/-$
& $\displaystyle{\frac{1}{8}\cdot
\frac{(2 J + 1)\:(2 J + 11)\:(2 J + 13)}
{(2 J + 5)\:(J + 2)}}$\\[10pt]
\enddata
\end{deluxetable}
\clearpage
\begin{deluxetable}{lr}
\tablenum{11}
\tablewidth{8.5cm}
\tablecaption{Estimated TiH Data for Thermochemistry}
\tablehead{
\colhead{State}  & \colhead{      Term value in cm$^{-1}$}}
\startdata
$X ^4\Phi_{3/2}$  &     0.0  \\
$X ^4\Phi_{5/2}$  &    99.6  \\
$X ^4\Phi_{7/2}$  &   199.2  \\
$X ^4\Phi_{9/2}$  &   298.8  \\
$1 ^4\Sigma^-$    &   970    \\
$1 ^4\Pi$         &  1732    \\
$1 ^2\Delta$      &  1770    \\
$1 ^4\Delta$      &  4231    \\
$1 ^2\Pi$         &  3390    \\
$1 ^2\Phi$        &  3710    \\
$1 ^2\Sigma^-$    &  5810    \\
$2 ^4\Sigma^-$    &  8630    \\
$2 ^4\Delta$      &  9520    \\
$2 ^4\Pi$         & 10810    \\
$A ^4\Phi$        & 10595    \\
$2 ^2\Pi$         &  6940    \\
$2 ^2\Delta$      & 13870    \\
$B ^4\Gamma$      & 18878    \\
\enddata
\label{table:X1}
\end{deluxetable}

\begin{deluxetable}{llllll}
\tablenum{12}
\tablewidth{10.5cm}
\tablecaption{Estimated Molecular Constants of TiH for Thermochemistry (in cm$^{-1}$)}
\tablehead{
\colhead{State}  & \colhead{$B_e$} &  \colhead{$\alpha_e$} & \colhead{$D_e$}  &
\colhead{$\omega_e$} & \colhead{$\omega_ex_e$}}
\startdata
$X ^4\Phi$     & 5.4975  &    0.286  &    2.56$\times10^{-4}$  &   1548.9  &    20.2 \\
$1 ^4\Sigma^-$ & 5.08    &    0.27   &    2.5$\times10^{-4}$   &   1484    &    22.8 \\
$1 ^4\Pi$      & 5.32    &    0.29   &    2.5$\times10^{-4}$   &   1487    &    20.4 \\
$1 ^2\Delta$   & 5.36    &    0.29   &    2.5$\times10^{-4}$   &   1519    &    26.9 \\
$1 ^4\Delta$   & 4.92    &    0.27   &    2.6$\times10^{-4}$   &   1399    &    20.6 \\
$1 ^2\Pi$      & 5.22    &    0.28   &    2.5$\times10^{-4}$   &   1450    &    24.6 \\
$1 ^2\Phi$     & 5.00    &    0.27   &    2.5$\times10^{-4}$   &   1462    &    26.6 \\
$1 ^2\Sigma^-$ & 5.02    &    0.27   &    2.5$\times10^{-4}$   &   1446    &    27.4 \\
$2 ^4\Sigma^-$ & 5.13    &    0.28   &    2.5$\times10^{-4}$   &   1469    &    27.4 \\
$2 ^4\Delta$   & 4.81    &    0.27   &    2.5$\times10^{-4}$   &   1367    &    23.2 \\
$2 ^4\Pi$      & 4.86    &    0.27   &    2.5$\times10^{-4}$   &   1403    &    29.6 \\
$A ^4\Phi$     & 4.9915  &    0.274  &    2.6$\times10^{-4}$   &   1342.6  &    21.3 \\
$2 ^2\Pi$      & 5.25    &    0.29   &    2.5$\times10^{-4}$   &   1564    &    24.7 \\
$2 ^2\Delta$   & 5.03    &    0.27   &    2.5$\times10^{-4}$   &   1435    &    22.1 \\
$B ^4\Gamma$   & 5.8906  &    0.303  &    3.6$\times10^{-4}$   &   1592.5  &    27.1 \\
\enddata
\label{table:X2}
\end{deluxetable}

\begin{deluxetable}{ll}
\tablenum{13}
\tablewidth{8.5cm}
\tablecaption{Derived Partition Function of TiH}
\tablehead{
\colhead{Temperature (K)}  & \colhead{Partition Function}}
\startdata
1200  &  1852   \\
1400  &  2560   \\
1600  &  3427   \\
1800  &  4469    \\
2000  &  5700   \\
2200  &  7136    \\
2400  &  8792    \\
2600  &  10682    \\
2800  &  12825    \\
3000  &  15236    \\
3200  &  17930    \\
3400  &  20926    \\
3600  &  24240    \\
3800  &  27890    \\
4000  &  31895    \\
4200  &  36272    \\
4400  &  41040    \\
4600  &  46217    \\
4800  &  51824    \\
\enddata
\label{table:X3}
\end{deluxetable}

\newpage


\end{document}